\documentclass{webofc}

\usepackage{graphicx}
\usepackage{bm}
\usepackage{epstopdf}

\woctitle{XLVII International Symposium on Multiparticle Dynamics}
\def\bea{\begin{eqnarray}}
\def\eea{\end{eqnarray}}

\def\pp{\mbox{$p$-$p$}}
\def\pa{\mbox{$p$-$A$}}

\def\auau{\mbox{Au-Au}}

\def\pbpb{\mbox{Pb-Pb}}
\def\aa{\mbox{$A$-$A$}}
\def\nn{\mbox{$N$-$N$}}

\def\ppbar{\mbox{$p$-$\bar p$}}

\def\pt{$p_t$}

\def\nch{$n_{ch}$}
\def\v2{$v_2$}
\def\mmpt{$\bar p_t$}
\def\ppb{\mbox{$p$-Pb}}
\def\pn{\mbox{$p$-$N$}}

\begin{document}
\title{Collectivity and manifestations of minimum-bias jets in high-energy nuclear collisions}

\author{Thomas A. Trainor\inst{1}\thanks{\email{ttrainor99@gmail.com}}
}

\institute{CENPA 354290 University of Washington, Seattle, Washington, USA
          }

\abstract{
Collectivity, as interpreted to mean flow of a dense medium in high-energy A-A collisions described by hydrodynamics, has been attributed to smaller collision systems -- p-A and even p-p collisions -- based on recent analysis of LHC data. However, alternative methods reveal that some data features attributed to flows are actually manifestations of minimum-bias (MB) jets. 
In this presentation I review the differential structure of single-particle $p_t$ spectra from SPS to LHC energies in the context of a two-component (soft + hard) model (TCM) of hadron production. I relate the spectrum hard component to measured properties of isolated jets. I use the spectrum TCM to predict accurately the systematics of ensemble-mean $\bar p_t$ in p-p, p-A and A-A collision systems over a large energy interval. 
Detailed comparisons of the TCM with spectrum and correlation data suggest that MB jets play a dominant role in hadron production near midrapidity. Claimed flow phenomena are better explained as jet manifestations agreeing quantitatively with measured jet properties.}
\maketitle

\section{Introduction} \label{intro}

Collectivity (interpreted to represent hydrodynamic flows~\cite{hydro}) has been attributed recently to smaller collision systems~\cite{dusling} based on certain correlation phenomena conventionally attributed to flows in larger \aa\ systems (e.g.\ $v_2$ data). But the term ``collectivity'' simply represents {\em any} correlated collection: correlations $\Rightarrow$ collectivity. For instance, dijet production represents a collective phenomenon that arguably dominates high-energy nuclear collisions. Several hadron production mechanisms may contribute to observed correlations. It should be our task to identify those mechanisms by comprehensive data analysis based on a variety of methods. The goal is then to extract {\em all}\, information from available particle data and interpret that information based on recognized physics principles. In this study measured  $\bar p_t$ trends for three collision systems are compared to corresponding \pt\ spectra and measured jet properties.

\section{Two-component (soft + hard) model or TCM for A-B collisions} \label{tcm}

The TCM is a comprehensive and accurate model of hadron production in A-B collisions near midrapidity inferred inductively from \pt\ spectra~\cite{ppprd,alicetomspec} and angular correlations~\cite{porter2,porter3,ppquad}. For instance, intensive charge density $\bar \rho_0$ can be separated into two components: $\bar \rho_0 = \bar \rho_s + \bar \rho_h$ (soft + hard). Based on comprehensive data analysis (e.g.~\cite{fragevo,ppquad,jetspec2}) the two components have been interpreted physically as follows: Soft component SC results from projectile-nucleon dissociation to {\em participant} low-$x$ gluons $\sim \bar \rho_s \equiv n_s / \Delta \eta \propto \log(\sqrt{s} / 10~ \text{GeV})$. Hard component HC results from large-angle scattering of participant gluons which then fragment to a minimum-bias (MB) ensemble of dijets. Measured SC and HC (yields, spectra and correlations) are related in \pp\ collisions via $\bar \rho_h \equiv n_h / \Delta \eta \approx \alpha \bar \rho_s^2$ ({\em noneikonal} trend) with $\alpha \approx O(0.01)$. An eikonal trend (e.g.\ as in the Glauber model of \aa\ collisions) would instead follow $\bar \rho_h \propto \bar \rho_s^{4/3}$~\cite{powerlaw}. The \pp\ $\bar \rho_h  \approx \alpha \bar \rho_s^2$ trend is illustrated in Fig.~\ref{jets} (a) as $n_h / n_s \approx 0.005\, \bar \rho_s$ (within $\Delta \eta = 1$)~\cite{ppprd}.

A TCM for hadron production in A-B collisions follows suite but requires additional geometry parameters defined within the Glauber model of collision geometry based on the eikonal approximation. $N_{part}/2$ is the number of participant nucleon pairs, $N_{bin}$ is the number of \nn\ binary collisions, and $\nu \equiv 2N_{bin} / N_{part}$ is the mean number of binary collisions per participant pair.
For instance, the event-ensemble mean of extensive total $P_t$  (\pt\ integrated within some acceptance $\Delta \eta$) can be represented in TCM form as $\bar P_t = \bar P_{ts} + \bar P_{th}$ or
\bea
&& \hspace{-.4in} \bar P_t =  (N_{part}/2) n_{sNN} \bar p_{tsNN} + N_{bin} n_{hNN} \bar p_{thNN};~~
\bar P_t / n_s = \bar p_{ts} + x(n_s) \nu(n_s) \bar p_{thNN}(n_s),
\eea
where $x \equiv \bar \rho_{hNN} / \bar \rho_{sNN} = n_{hNN} / n_{sNN}$ and the TCM incorporates factorization of N-N low-$x$ gluon participants (noneikonal) and A-B nucleon participants (eikonal). Note that \mmpt\ SC $\bar p_{tsNN} \rightarrow \bar p_{ts} \approx 0.4$ GeV/c is observed to be universal for \nn\ collisions within A-B collisions.


\section{Minimum-bias jet manifestations in p-p $\bf p_t$ spectra and correlations} \label{}

Quantitative correspondence between measured jet properties and single-particle \pt\ spectra has been demonstrated previously for 200 GeV \pp\ and \auau\ spectra down to 0.5 GeV/c~\cite{fragevo}.  The demonstration relies on the observation that jet spectra and  fragmentation functions (FFs) have simple universal forms (Gaussian and beta distribution) when plotted on rapidity variables. Some details are reviewed in this section (angular correlations are described below).

 \begin{figure}[h]
\centering
  \includegraphics[height=.25\textwidth]{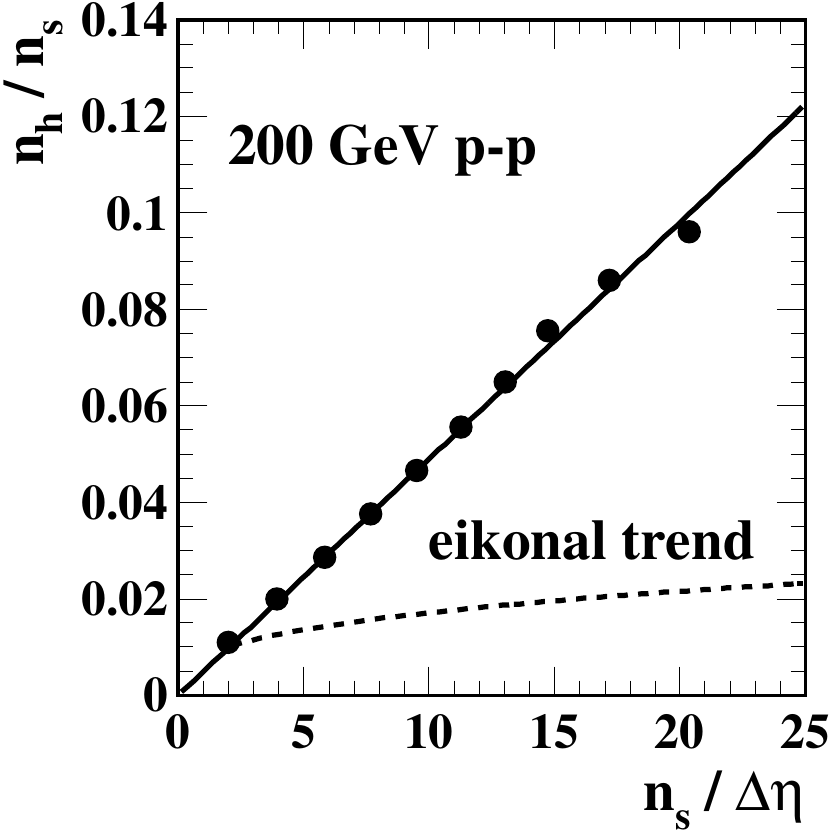}
\put(-70,65) {\bf (a)}
  \includegraphics[height=.25\textwidth]{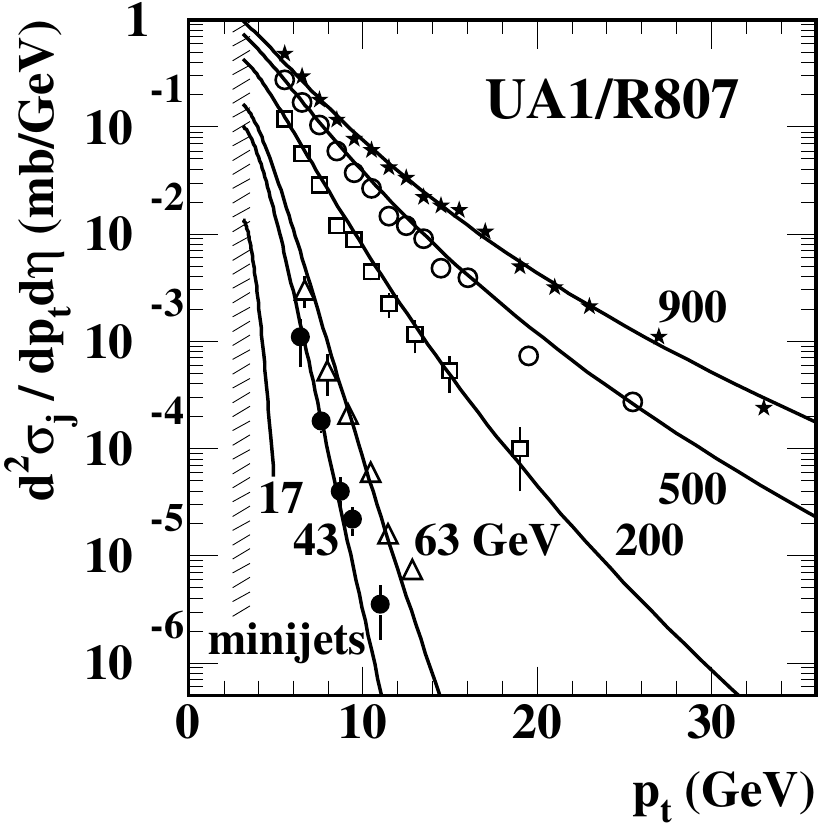}
\put(-23,70) {\bf (b)}
  \includegraphics[height=.25\textwidth]{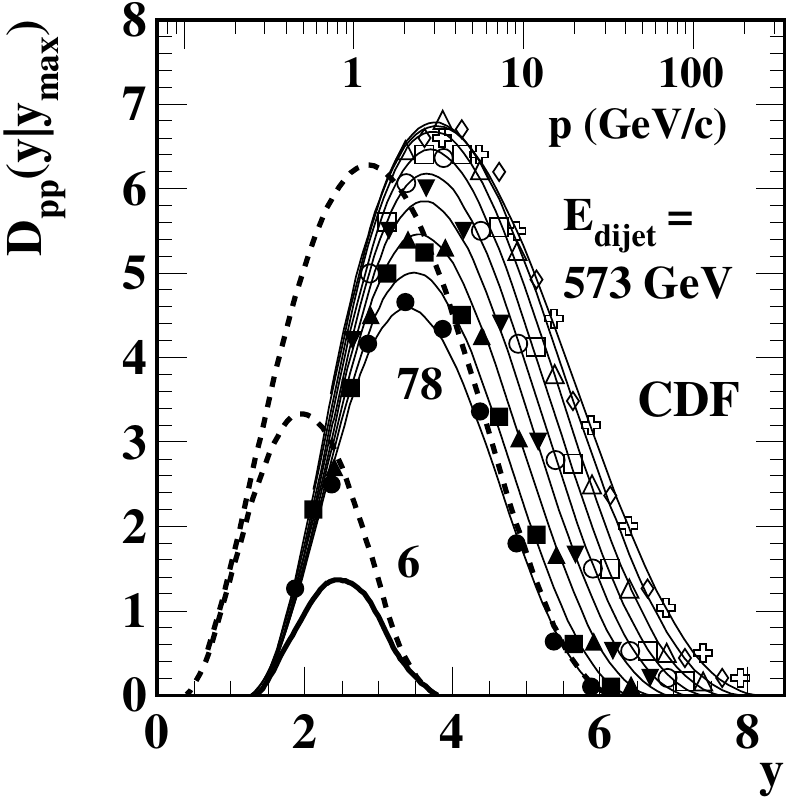}
\put(-70,78) {\bf (c)}
  \includegraphics[height=.25\textwidth]{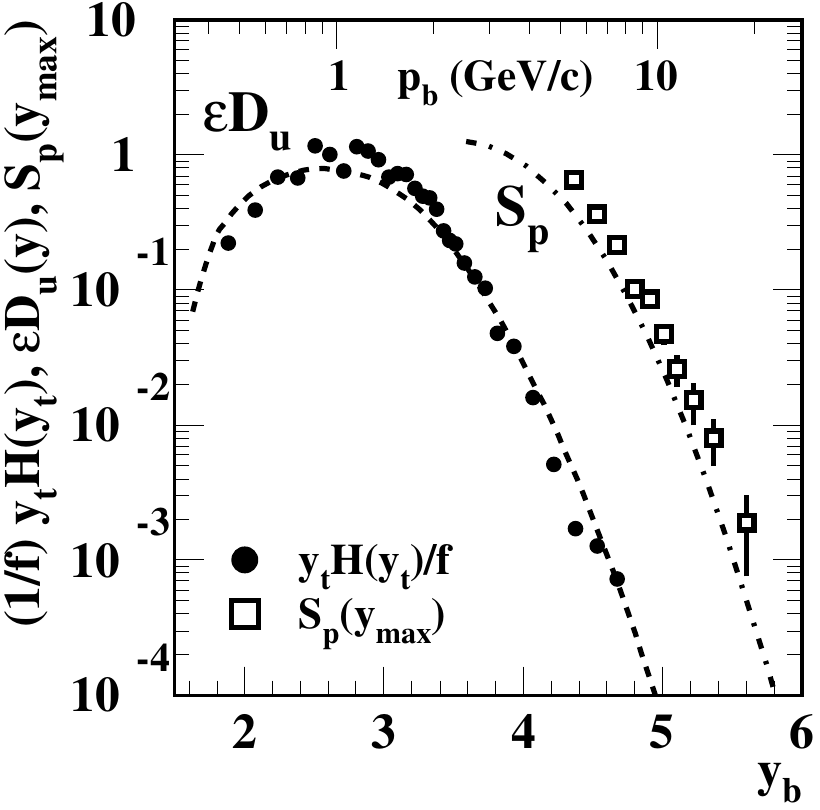}
\put(-23,76) {\bf (d)}
\caption{\label{jets}
(a) Hard/soft ratio vs soft density $\bar \rho_s$.
(b) Jet \pt\ spectra for several energies from the ISR and Sp\=pS. Curves through data are from Ref.~\cite{jetspec2}.
(c) Fragmentation functions from \ppbar\ collisions described by a simple parametrization from Ref.~\cite{jetspec2}.
(d) Spectrum hard component for 200 GeV NSD \pp\ collisions (solid points) compared to a fragment distribution as predicted by Eq.~(\ref{convolute}) (dashed).
} 
 \end{figure}

Figure~\ref{jets} (b) shows jet spectra from the ISR (43 and 63 GeV points) and Sp\=pS (remaining data points). The curves are derived from a universal parametrization of jet spectra applicable up to 13 TeV~\cite{jetspec2}. Panel (c) shows jet FFs from \ppbar\ collisions obtained by the CDF collaboration plotted on rapidity variable $y = \ln[(E + p)/m_\pi]$. The FFs show self-similar variation with jet energy making a simple parametrization possible. Schematically, the {\em fragment distribution} (FD) $P(p)$ describing the contribution to a \pt\ spectrum from MB dijets (spectrum HC) is the convolution of jet spectrum $P(E)$ for a given \pp\ collision energy and FF ensemble $P(p|E)$
\bea \label{convolute}
P(p) &=& \int_{E_{min}}^\infty dE\, P(p|E) \, P(E),
\eea
with $E_{min} \approx 3$ GeV established by comparisons with hadron spectra~\cite{fragevo,pptrig}. In panel (d) the dashed curve is $P(d)$. The solid points are the spectrum HC from 200 GeV NSD \pp\ collisions~\cite{ppprd}. The open boxes are the 200 GeV UA1 jet spectrum appearing in panel (b). It is notable  that the majority of jet fragments appear near 1 GeV/c. If $E_{min}$ is reduced significantly below 3 GeV a large overestimate of the measured FD contribution below 1 GeV/c results, since the jet spectrum scales as $d\sigma_{jet}/ dp_t \sim 1/p_t^6$ near that energy.

Two-particle angular correlations have been studied for 200 GeV \pp\ collisions~\cite{porter2,porter3,ppquad} and \auau\ collisions~\cite{anomalous} within $\Delta \eta = 2$. Measured quantity $\Delta \rho / \sqrt{\rho_{ref}}$ is essentially the number of correlated pairs per final-state hadron; $\eta_\Delta$ and $\phi_\Delta$ are angle difference variables.

 \begin{figure}[h]
\centering
  \includegraphics[width=.25\textwidth,height=.25\textwidth]{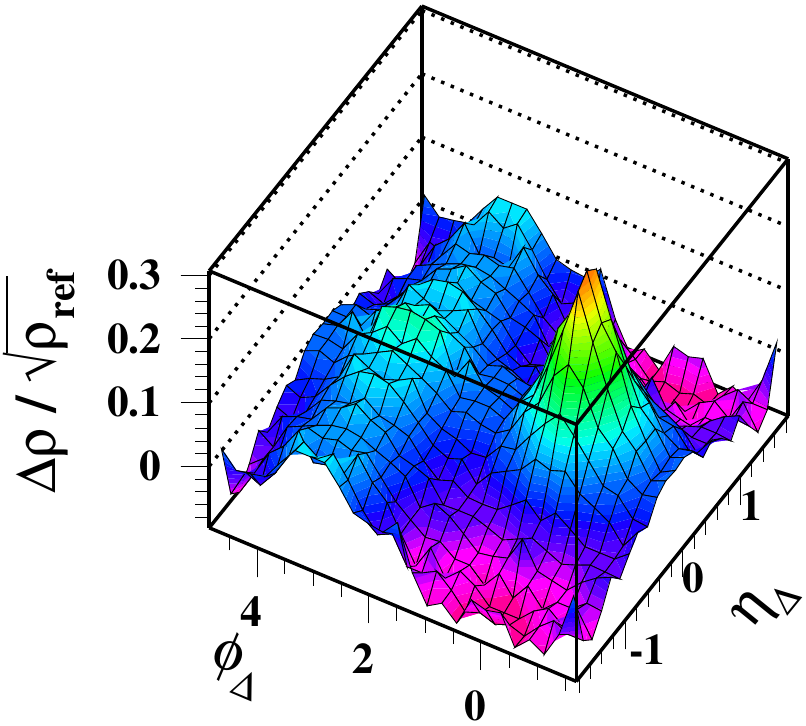}
\put(-80,86) {\bf (a)}
  \includegraphics[height=.25\textwidth]{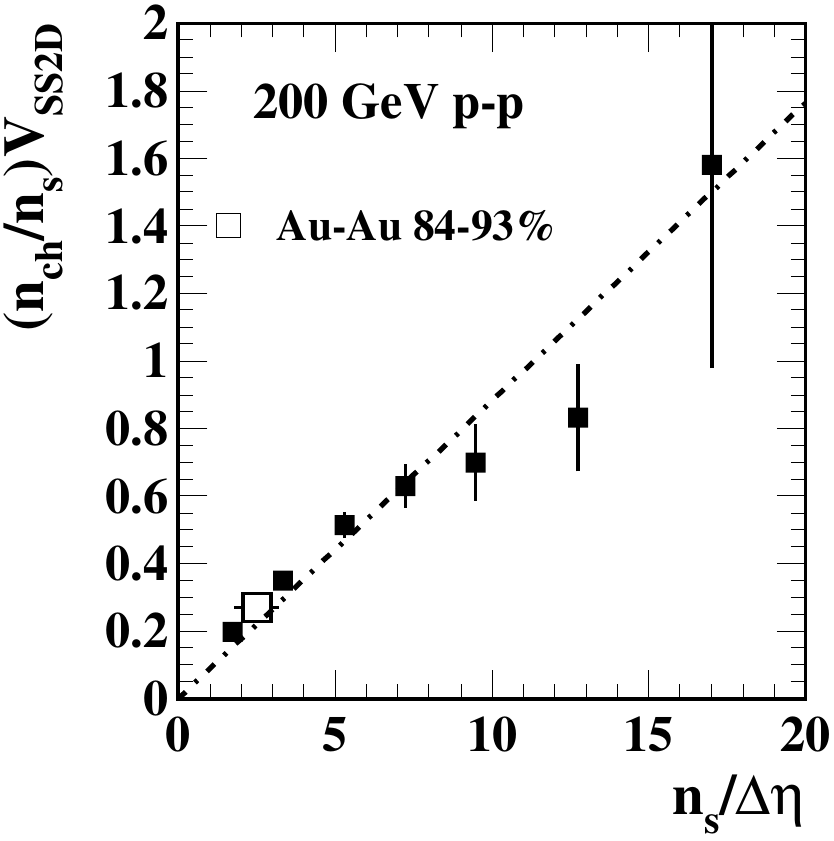}
\put(-70,50) {\bf (b)}
  \includegraphics[height=.25\textwidth]{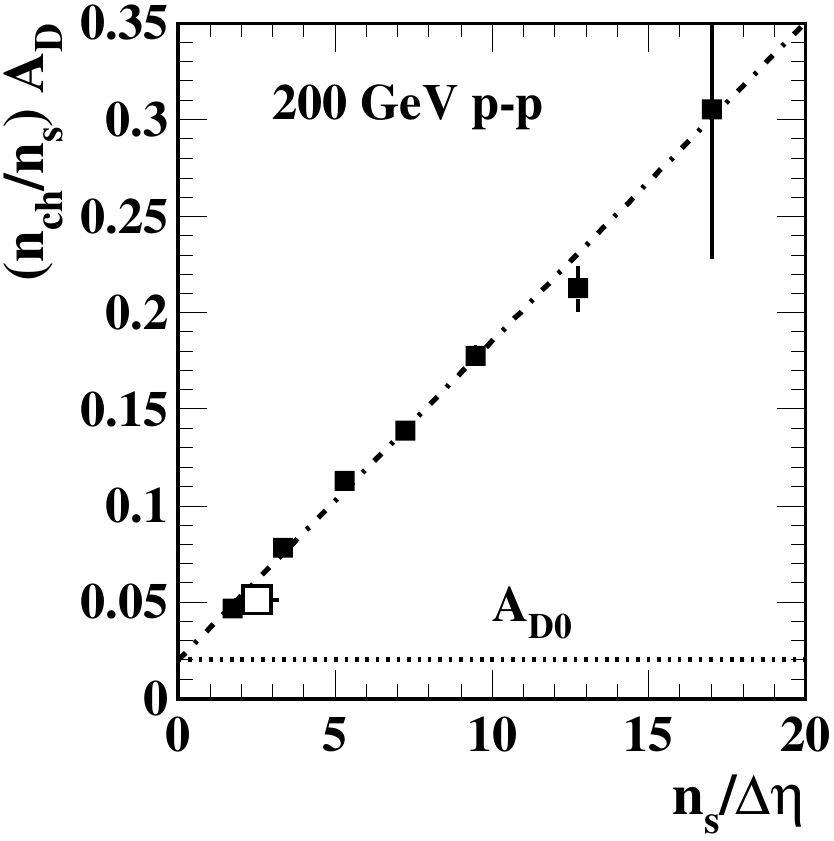}
\put(-70,72) {\bf (c)}
  \includegraphics[height=.25\textwidth]{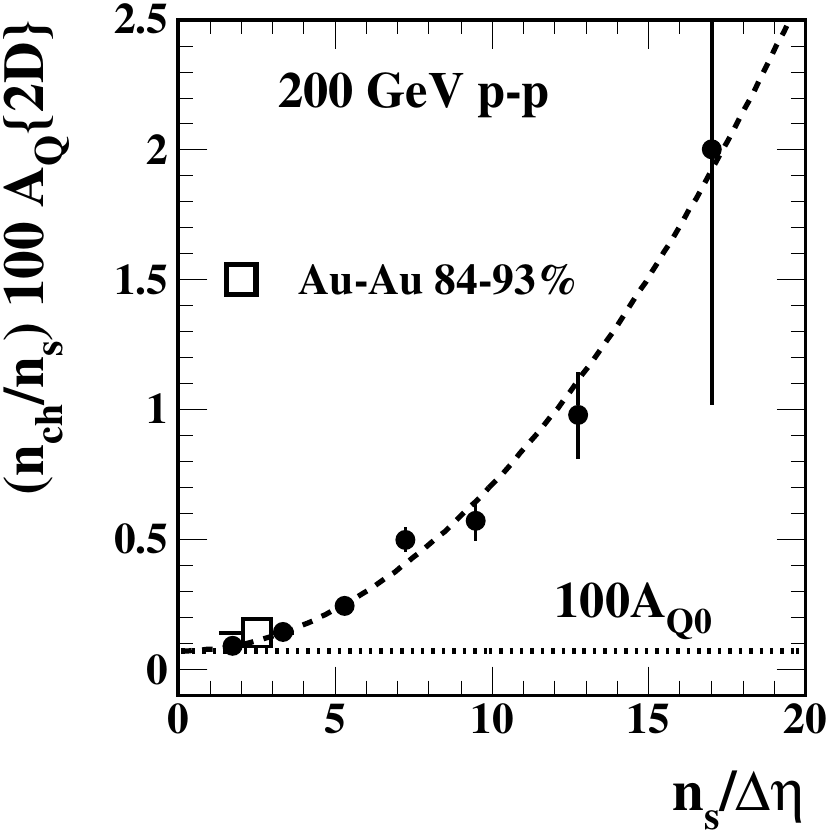}
\put(-70,72) {\bf (d)}
\caption{\label{angcorr}
(a) 2D angular correlations for high-multiplicity 200 GeV \pp\ collisions.
(b) SS 2D peak volume vs soft density $\bar \rho_s$.
(c) AS 1D peak amplitude vs density $\bar \rho_s$.
(d) Quadrupole amplitude $A_Q$ vs $\bar \rho_s$. In (b-d) the vertical axis is proportional to number of correlated pairs divided by $\bar \rho_s$.
} 
 \end{figure}

Figure~\ref{angcorr} (a) shows angular correlations for high-multiplicity ($\approx 8 \bar \rho_{0NSD}$) 200 GeV \pp\ collisions with no special \pt\ cuts. The main features are a SS (same-side on azimuth) 2D peak representing individual jets and an AS  (away-side) 1D peak on $\phi_\Delta$ representing back-to-back jet pairs. Superposed on the broader SS peak is a narrower 2D contribution from Bose-Einstein correlations. Just visible is a small  SC contribution -- a narrow 1D peak on $\eta_\Delta$.

Figure~\ref{angcorr} (b-d) shows systematic variation with \nch\ or $n_s$ of individual correlation amplitudes obtained by 2D model fits to correlation data~\cite{ppquad}. The linear per-hadron trends in panels (b,c) indicate that the correlation hard component (dijets) follows a noneikonal trend: jet-correlated pairs $\propto \bar \rho_s^2$ as observed for the \pp\ spectrum HC. 
Panel (d) shows a quadratic trend on $\bar \rho_s$ for {\em nonjet quadrupole} component $A_Q$ indicating that quadrupole correlated pairs are $\propto \bar \rho_s^3$. That trend is actually consistent with the quadrupole trend observed for \aa\ collisions. If $\bar \rho_s \sim$ low-$x$ participant gluons and $\bar \rho_s^2 \sim$ participant-gluon binary collisions then $\bar \rho_s^3 \sim N_{part} N_{bin}$ for \pp\ collisions. The centrality trend observed for the nonjet quadrupole ($\bar \rho_0^2 v_2^2$) in \auau\ collisions is $N_{part} N_{bin} \epsilon_{opt}^2$~\cite{davidhq}. Absence of a $\epsilon_{opt}^2$ factor for \pp\ collisions is consistent with the noneikonal $\bar \rho_h \propto \bar \rho_s^2$ trend: centrality is not relevant for \pp\ collisions.


\section{p-p $\bf p_t$ spectrum TCM hard components} \label{}

The illustrations above pertain to 200 GeV RHIC data as a testbed for MB dijet manifestations in spectra and angular correlations. In preparation for \mmpt\ analysis the full collision-energy and charge-multiplicity dependence of the spectrum hard component should be understood.

\begin{figure}[h]
\centering
  \includegraphics[height=.245\textwidth]{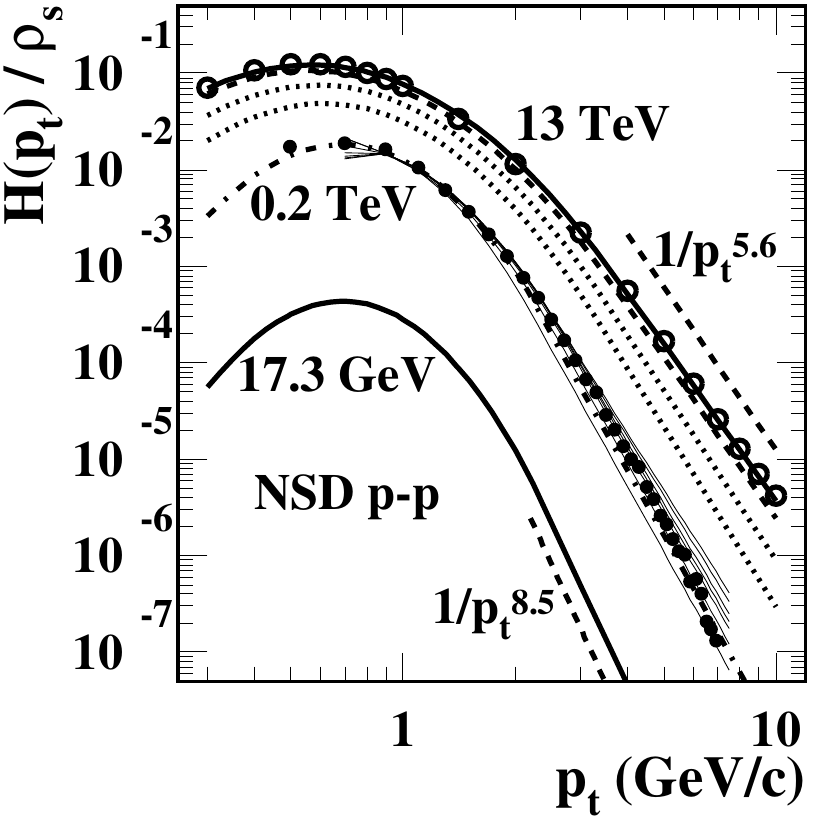}
\put(-18,86) {\bf (a)}
  \includegraphics[height=.25\textwidth]{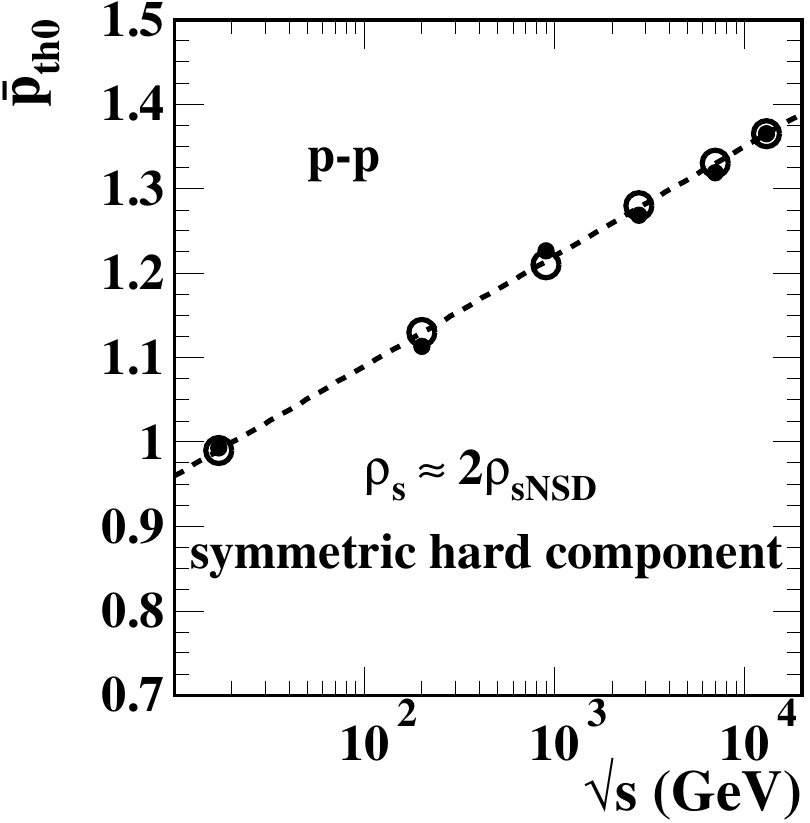}
\put(-70,84) {\bf (b)}
  \includegraphics[height=.245\textwidth]{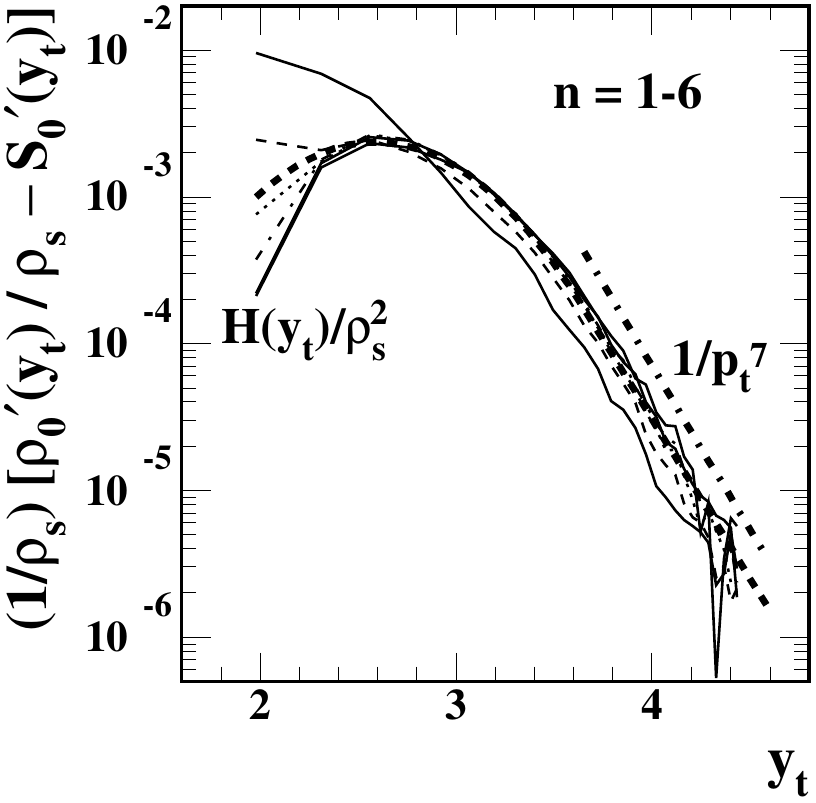}
\put(-70,26) {\bf (c)}
  \includegraphics[height=.25\textwidth]{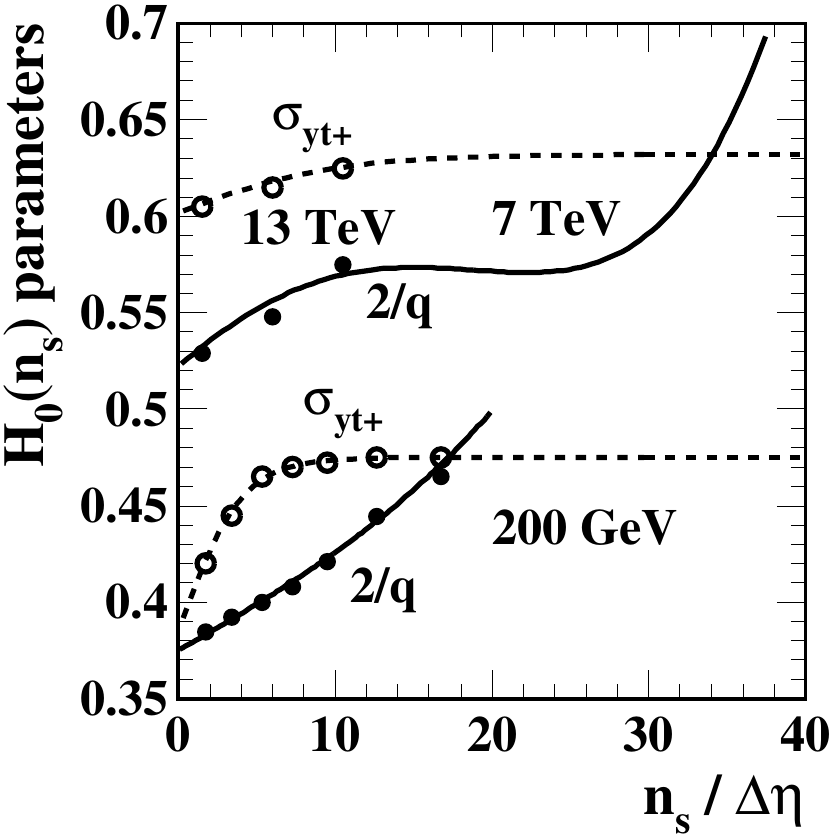}
\put(-24,26) {\bf (d)}
\caption{\label{ppspectra}
(a) \pp\ \pt\ spectrum TCM hard components for several collision energies (curves) compared to data (points).
(b) Mean-\pt\ \mmpt\ hard components $\bar p_{th0}$ inferred from TCM model functions in panel (a).
(c) \pt\ spectrum hard components for several multiplicity classes of 200 GeV \pp\ collisions.
(d) Model parameters $\sigma_{y_t+}$ and $q$ for the TCM spectrum hard component, their variation with \nch.
} 
 \end{figure}

Figure~\ref{ppspectra} (a) shows the TCM hard components for \pp\ \pt\ spectra as a function of collision energy from $\sqrt{s} = 17$ GeV to 13 TeV (curves of various line styles)~\cite{alicetomspec} compared to data for two energies (points)~\cite{ppprd,alicespec}. The TCM spectrum HC model is simple and relies on a few $\log(\sqrt{s})$ terms. The description of spectrum data is in all cases within systematic uncertainties of the data. Spectrum hard components show significant dependence on event multiplicity \nch\ (e.g.\ multiple fine lines for the 200 GeV model). For the \mmpt\ study fiducial hard-component $\bar p_{th0}$ values in panel (b) are inferred from a {\em symmetric} spectrum HC model (for $\bar \rho_s \approx 2\bar \rho_{sNSD}$)~\cite{alicetommpt}.

Figure~\ref{ppspectra} (c) shows spectrum hard components for several \nch\ classes of 200 GeV \pp\ collisions~\cite{ppquad}. The \nch\ dependence of the shape has been known for more than ten years~\cite{ppprd} and is attributed to bias of the underlying jet energy spectrum in response to the \nch\ condition. The TCM is modified to accommodate those changes as follows: The hard-component model is a Gaussian plus exponential tail. The transition from Gaussian to exponential is determined by slope matching. The Gaussian widths are treated separately below and above  the mode with widths $\sigma_{y_t-}$ and $\sigma_{y_t+}$ respectively. The exponential constant $q$ is a separate parameter.

Figure~\ref{ppspectra} (d) shows evolution of spectrum HC model parameters $\sigma_{y_t+}$ and $q$ with \nch\ at 200 GeV and 13 TeV. The trend for the width below the mode $\sigma_{y_t-}$ is available in Ref.~\cite{alicetomspec}. For 200 GeV the parameter trends are tightly constrained by spectrum data over the full \nch\ interval relevant to \mmpt\ data. For 13 TeV the spectrum data span a more-limited \nch\ interval. The $2/q$ trend (upper solid curve) has been extended by fitting \mmpt\ data in the present study.

The \mmpt\ analysis of Ref.~\cite{alicempt} produced uncorrected $\bar p_t'$ values, where the prime denotes the effect on spectra of a lower-\pt\ acceptance cut at $p_{t,min}$ near 0.15 GeV/c. The \pt\ acceptance cut strongly affects the \mmpt\ SC but has negligible effect on the HC which is small below 0.5 GeV/c. The relation is $\bar p_{ts}' = \bar p_{ts} / \xi$ where $\xi$ is the SC efficiency $\approx 0.78$ for an {\em effective} \pt\ cutoff near 0.17 GeV/c. For universal $\bar p_{ts} \approx 0.4$ GeV/c uncorrected $\bar p_{ts}' \approx 0.51\pm0.2$ GeV/c.


\section{p-p $\bf \bar p_t$ TCM analysis} \label{pp}

The TCM for \mmpt\ data from \pp\ collisions is constructed as follows  \cite{alicetommpt}: The uncorrected yield $n_{ch}'$ within angular acceptance $\Delta \eta$ relative  to the corrected soft component $n_s$ is $n_{ch}' / n_s = \xi + x(n_s)$ with $\xi = 0.76$ - 0.80 (depending on the effective $p_{t,min}$). For noneikonal \pp\ collisions $x(n_s,\sqrt{s}) \equiv n_h / n_s \approx \alpha(\sqrt{s}) \bar \rho_s$ and $\alpha(\sqrt{s})$ is predicted from jet characteristics~\cite{alicetomspec}. The TCM for uncorrected $\bar p_t'$ is the first of
\bea \label{pptcm}
&& \hspace{-.4in} \bar p_t' \equiv \frac{\bar P_t'}{n_{ch}'} \approx \frac{\bar p_{ts} + x(n_s,\sqrt{s})\, \bar p_{th}(n_s,\sqrt{s})}{\xi + x(n_s,\sqrt{s})};~~~  \frac{n_{ch}'}{n_s} \bar p_t'  \approx \frac{\bar P_t}{n_s} = \bar p_{ts} + x(n_s,\sqrt{s})\, \bar p_{th}(n_s,\sqrt{s}),
\eea
where $\bar p_{ts} \approx 0.4$ GeV/c and $p_{th}(n_s,\sqrt{s})$ is the new information derived from \pp\ \mmpt\ data.

 \begin{figure}[h]
\centering
  \includegraphics[height=.25\textwidth]{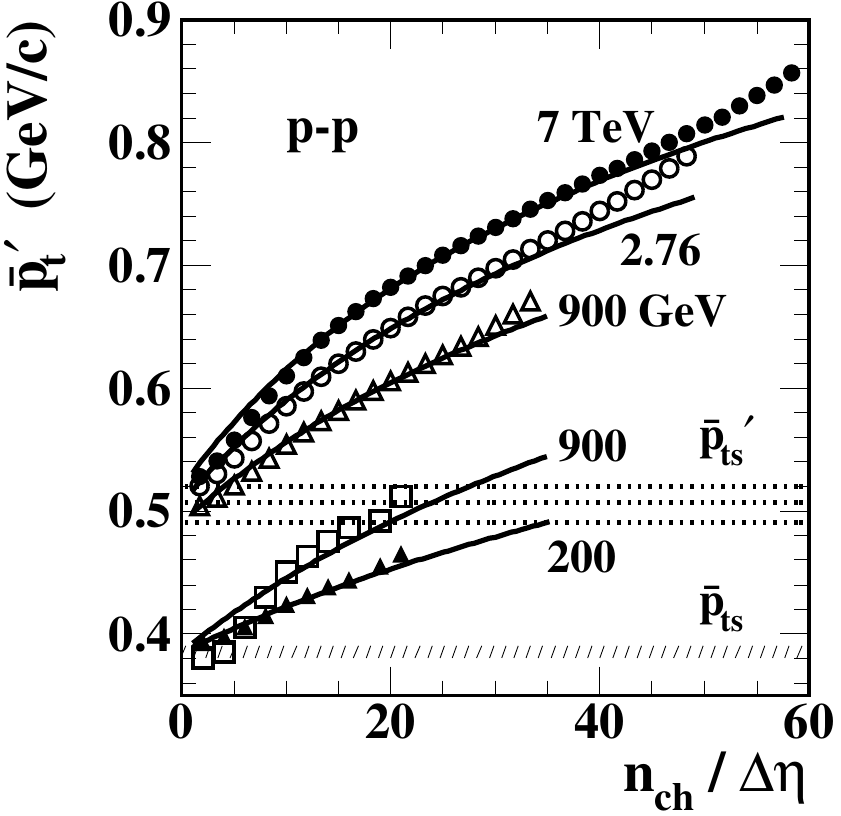}
\put(-70,66) {\bf (a)}
  \includegraphics[height=.25\textwidth]{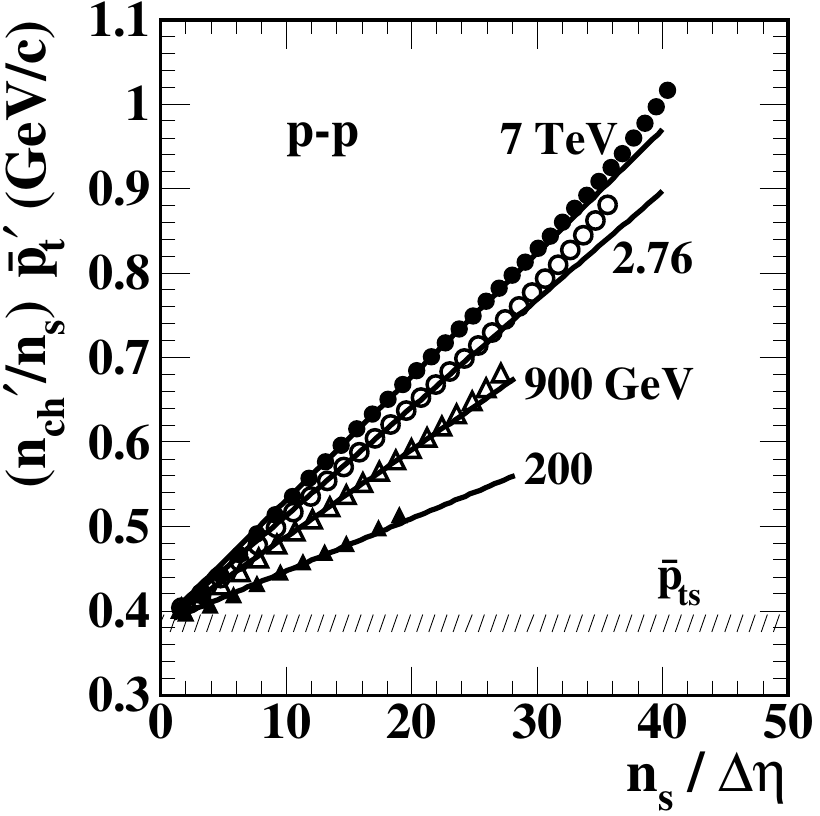}
\put(-70,66) {\bf (b)}
  \includegraphics[height=.25\textwidth]{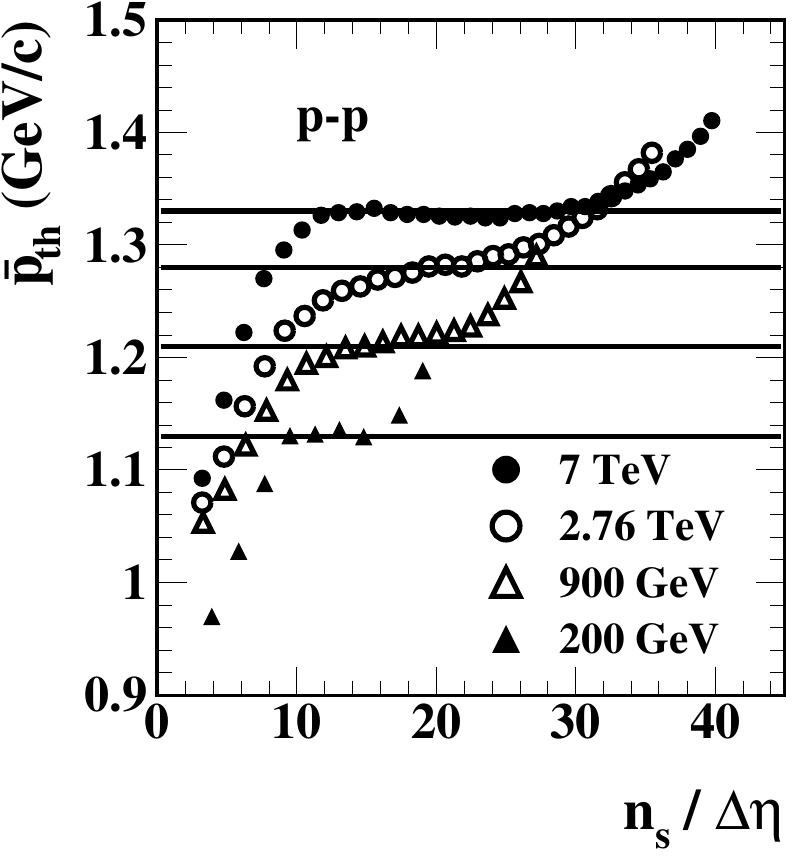}
\put(-60,30) {\bf (c)}
  \includegraphics[height=.25\textwidth]{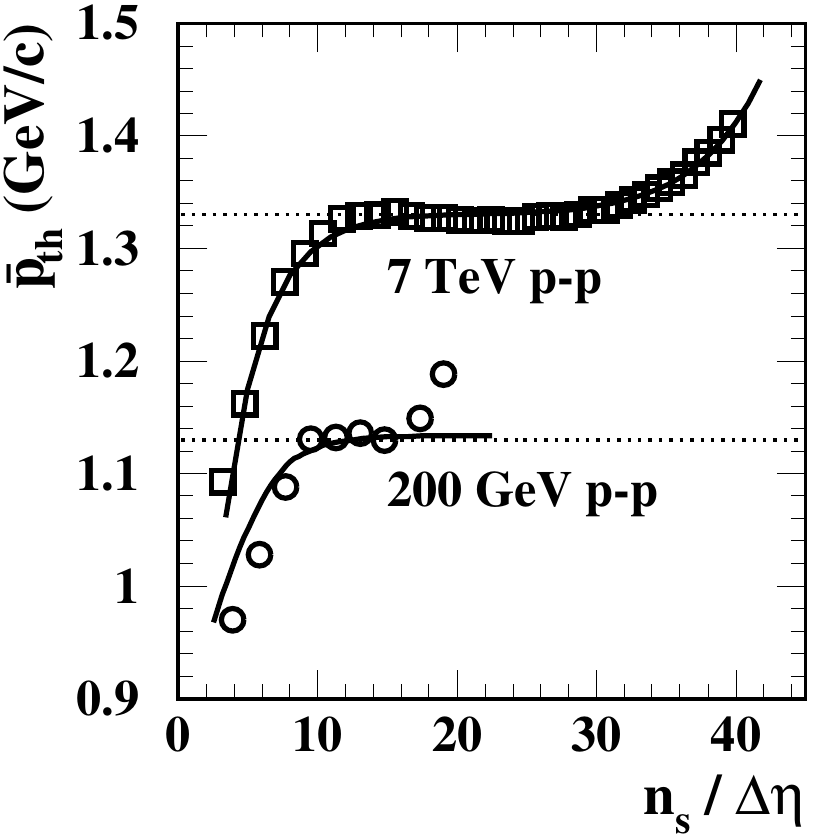}
\put(-60,30) {\bf (d)}
\caption{\label{ppmpt}
(a) Uncorrected $\bar p_{t}'$ data (points) from Ref.~\cite{alicempt} vs the \mmpt\ TCM (solid curves).
(b) Corrected $\bar P_{t} / n_s$ data (points) vs the TCM (solid lines).
(c) $\bar p_{th}(n_s,\sqrt{s}),$ trends vs $\bar \rho_s$ for four energies.
(d) $\bar p_{th}(n_s,\sqrt{s}),$ trends for 200 GeV and 7 TeV (points) compared to TCM trends from spectra (curves).
} 
 \end{figure}

Figure~\ref{ppmpt} (a) shows uncorrected $\bar p_t'$ data from Ref.~\cite{alicempt} (upper points) compared to nominally corrected data from STAR and UA1 (lower points). The $\bar p_{ts}'$ values for uncorrected data are in the range 0.49 - 0.525 GeV/c as expected (dotted lines). The solid curves are TCM trends assuming no \nch\ variation of \mmpt\ hard component [held fixed at values $\bar p_{th0}$ from Fig.~\ref{ppspectra} (b)].
Figure~\ref{ppmpt} (b) shows corrected $\bar P_t / n_s$ data (points) according to the second of Eqs.~(\ref{pptcm}). In that format the TCM is a straight line with slope $\alpha(\sqrt{s})\, \bar p_{th0}$ predicted from \pt\ spectra and jet data.

Figure~\ref{ppmpt} (c) shows $\bar p_{th}(n_s,\sqrt{s})$ data inferred from the uncorrected $\bar p_{t}'$ data according to Eqs.~(\ref{pptcm}). The horizontal lines are the $\bar p_{th0}(\sqrt{s})$ values appearing in Fig.~\ref{ppspectra} (b). Clearly evident are substantial variations with $n_s$ or $\bar \rho_s$. However, such variations are expected based on previous spectrum studies as in Refs.~\cite{ppprd,alicetomspec}. Figure~\ref{ppmpt} (d) shows $\bar p_{th}(n_s)$ values for 200 GeV and 7 TeV (points) compared to a TCM including HC \nch\ dependence according to Fig.~\ref{ppspectra} (d). The correspondence between TCM and \mmpt\ data is within data uncertainties. Accurate correspondence between \pp\ \mmpt\ data and the TCM compels the conclusion that  there is a direct, quantitative correspondence among measured isolated-jet properties, \pt\ spectrum hard components and \mmpt\ data. \pp\ collisions are dominated by MB dijets and are noneikonal.


\section{p-Pb $\bf \bar p_t$ TCM analysis} \label{ppb}

For a TCM description of \ppb\ collisions  \cite{alicetommpt} the composite structure of the Pb nucleus must be incorporated by adding {\em mean participant pathlength} $\nu(n_s) \equiv 2 N_{bin} / N_{part}$ as a geometry measure, so for instance $n_{ch}' / n_s \rightarrow \xi + x(n_s) \nu(n_s)$. The soft-component charge density factorizes as $\bar \rho_s \equiv (N_{part}/2) \bar \rho_{sNN}(n_s)$, similarly $\bar \rho_h \equiv N_{bin}\, \bar \rho_{hNN}(n_s)$ and hard/soft ratio $x(n_s) \rightarrow \bar \rho_{hNN}(n_s) / \bar \rho_{sNN}(n_s) \approx \alpha  \bar \rho_{sNN}(n_s)$ averaged over \nn\ collisions as for \pp\ collisions. Combining those relations gives $N_{part}(n_s)/2 = \alpha \bar \rho_s / x(n_s)$, and for \pa\ collisions $N_{part} \equiv N_{bin} + 1$. Thus, if a model for $x(n_s)$ is specified then $N_{part}$, $N_{bin}$ and $\nu$ are also defined. As described below, $x(n_s)$  for more-peripheral \pa\ collisions is based on the noneikonal \pp\ trend and is then extrapolated to more-central collisions with $N_{part}/2 > 1$ based on \mmpt\ data.
The TCM for \ppb\ $\bar p_t'$ data is similar to Eqs.~(\ref{pptcm}) but with additional $\nu(n_s)$ factors
\bea \label{ppbtcm}
\bar p_t' \equiv \frac{\bar P_t'}{n_{ch}'} \approx \frac{\bar p_{ts} + x(n_s) \nu(n_s)\, \bar p_{thNN}(n_s)}{\xi + x(n_s) \nu(n_s)};~~~  \frac{\bar P_t(n_s)}{n_s} = \bar p_{ts} + x(n_s) \nu(n_s)\, \bar p_{thNN}(n_s)~
\eea
assuming $\bar p_{thNN}(n_s) \approx \bar p_{th0}$ (\pp\ values) for \ppb\ collisions (no jet modification in \pa).

 \begin{figure}[h]
\centering
  \includegraphics[height=.25\textwidth]{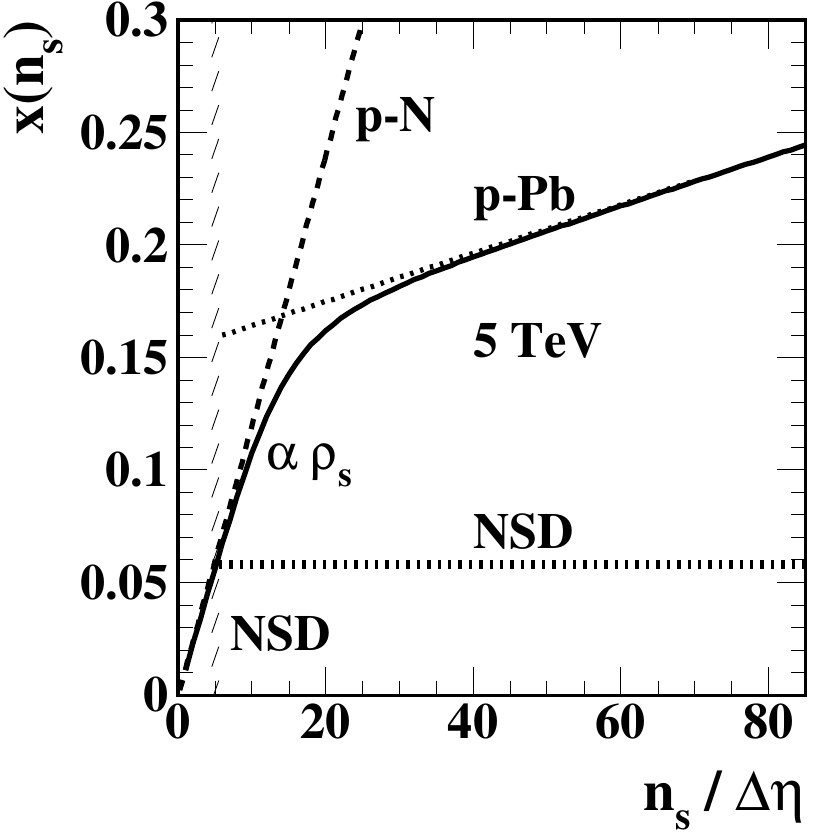}
\put(-20,55) {\bf (a)}
  \includegraphics[height=.25\textwidth]{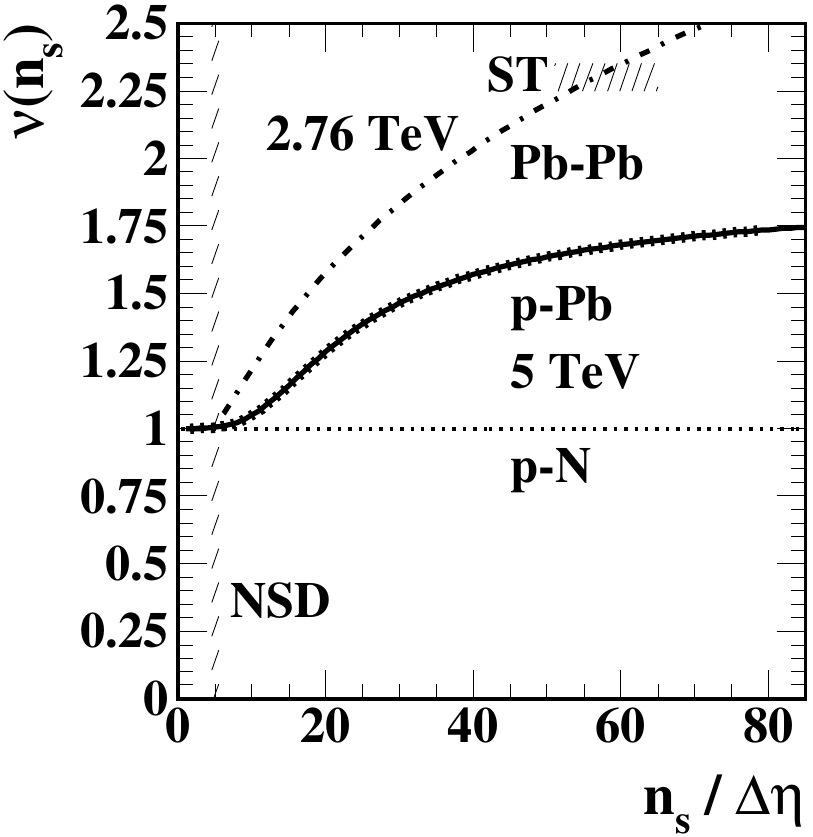}
\put(-19,55) {\bf (b)}
  \includegraphics[height=.25\textwidth]{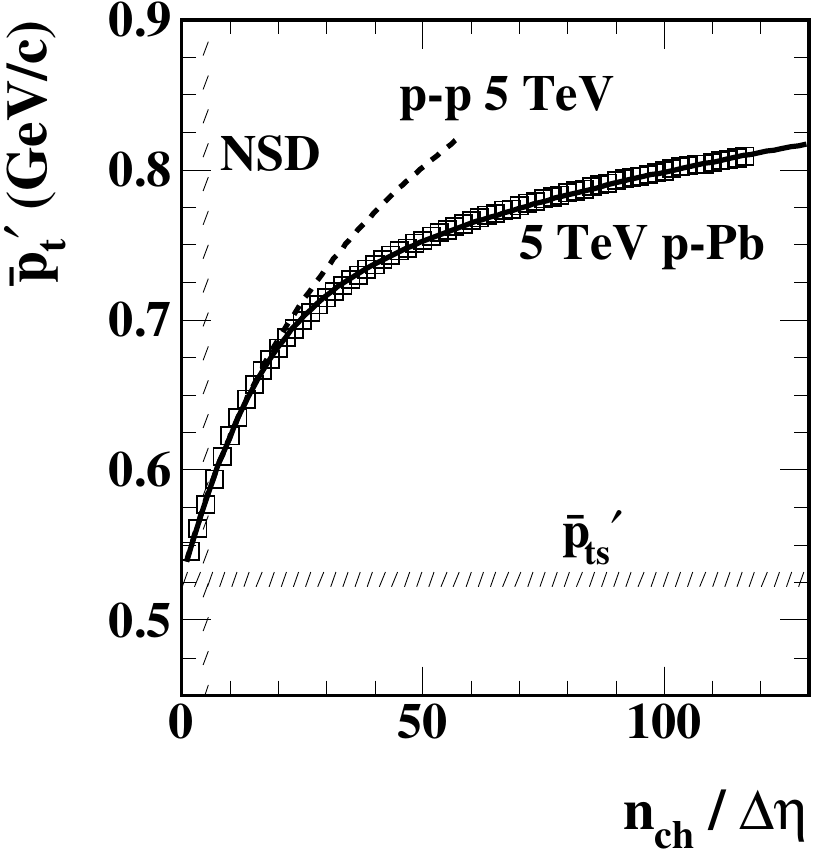}
\put(-20,55) {\bf (c)}
  \includegraphics[height=.25\textwidth]{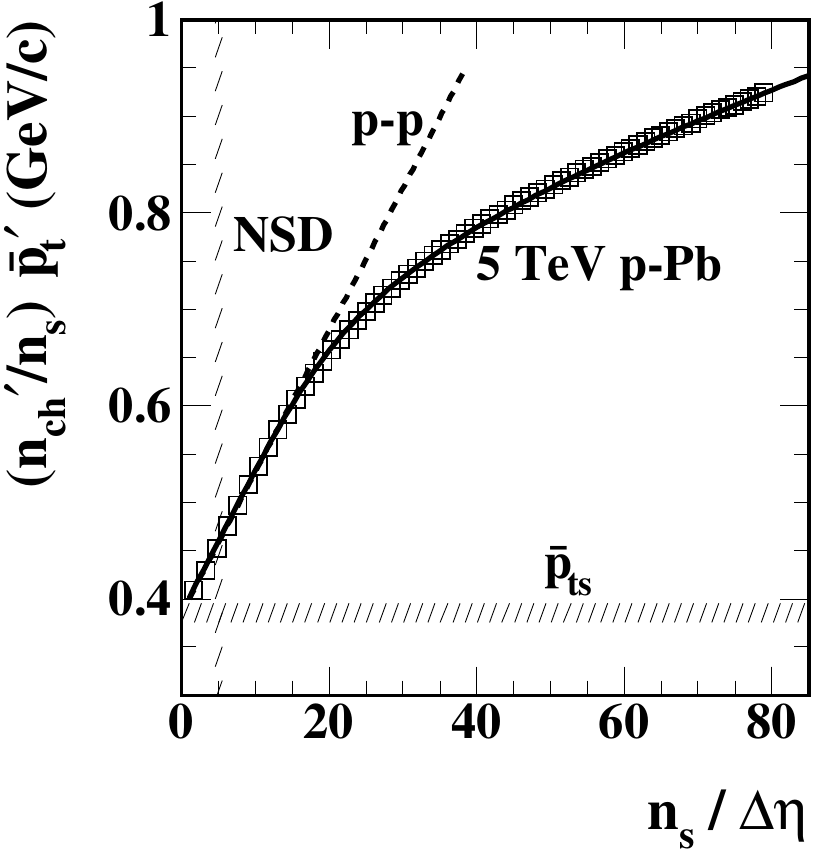}
\put(-20,55) {\bf (d)}
\caption{\label{ppbmpt}
(a) TCM parameter $x(n_s)$ vs $\bar \rho_s$ for 5 TeV \ppb\ collisions.
(b) Corresponding $\nu(n_s)$ trend.
(c) Uncorrected $\bar p_{t}'$ data (points) from Ref.~\cite{alicempt} vs the \mmpt\ TCM (solid curve).
(d) Corrected $\bar P_{t} / n_s$ data (points) vs the TCM (solid curve). Equivalent \pp\ trends are included for comparison.
} 
 \end{figure}

Figure~\ref{ppbmpt} (a) shows a $x(n_s)$ model for \ppb\ collisions (solid curve): the observed \pp\ trend (dashed line) is smoothly transitioned to a second linear trend with reduced slope (dotted line). The additional \ppb\ model parameters are $(\bar \rho_{s0},m_0)$, a transition point and slope-reduction factor. Panel (b) shows the corresponding $\nu(n_s)$ trend (solid curve) defined by $x(n_s)$ as noted above. The dash-dotted curve is a \pbpb\ $\nu(n_s)$ trend for comparison.

Figure~\ref{ppbmpt} (c) shows uncorrected $\bar p_t'$ data from Ref.~\cite{alicempt} (points) with $\bar p_{ts}' \approx 0.525$ GeV/c. The solid curve is the TCM defined by Eq.~(\ref{ppbtcm}) (first) with $\bar \rho_{s0} = 15$ and $m_0 = 0.1$ adjusted to accommodate  $\bar p_t'$ data and $\bar p_{thNN}(n_s) \rightarrow \bar p_{th0} = 1.30$ GeV/c from Fig.~\ref{ppspectra} (b). The dashed curve is the \pp\ TCM for 5 TeV. Figure~\ref{ppbmpt} (d) shows the same features for corrected $\bar P_t / n_s$ data with TCM described by Eq.~(\ref{ppbtcm}) (second). This \ppb\ exercise illustrates the transition from noneikonal \pp\ or \pn\ trend to eikonal \pa\ trend with increasing multiplicity. The value of $N_{part} / 2$ remains near one (\pn\ only) up to the  transition point but then slowly increases as greater \pa\ centrality becomes competitive to deliver larger overall charge multiplicity.


\section{Pb-Pb $\bf \bar p_t$ TCM analysis} \label{pbpb}

For the \pbpb\ \mmpt\ TCM published yield measurements and a Glauber model based on the eikonal approximation can be used to determined some TCM model elements. Parameter $x(n_s)$ can be derived from {\em per-participant-pair} charge-density data based on the TCM
\bea \label{nchtcm}
(2 / N_{part}) dn_{ch}/d\eta \approx  (2 / N_{part})\bar \rho_0(\nu) &=& \bar \rho_{sNN} [1 + x(\nu) \nu].
\eea
Parameter $\nu(n_s)$ can be derived from the Glauber model in the form $\nu(n_s) \approx (N_{part}/2)^{1/3}$~\cite{powerlaw} with $N_{part} / 2 \approx \bar \rho_{s} / \bar \rho_{sNN}$ as in the previous section and $\bar \rho_{sNN} \approx \bar \rho_{sNSD}$ in both cases.

\begin{figure}[h]
\centering
  \includegraphics[height=.25\textwidth]{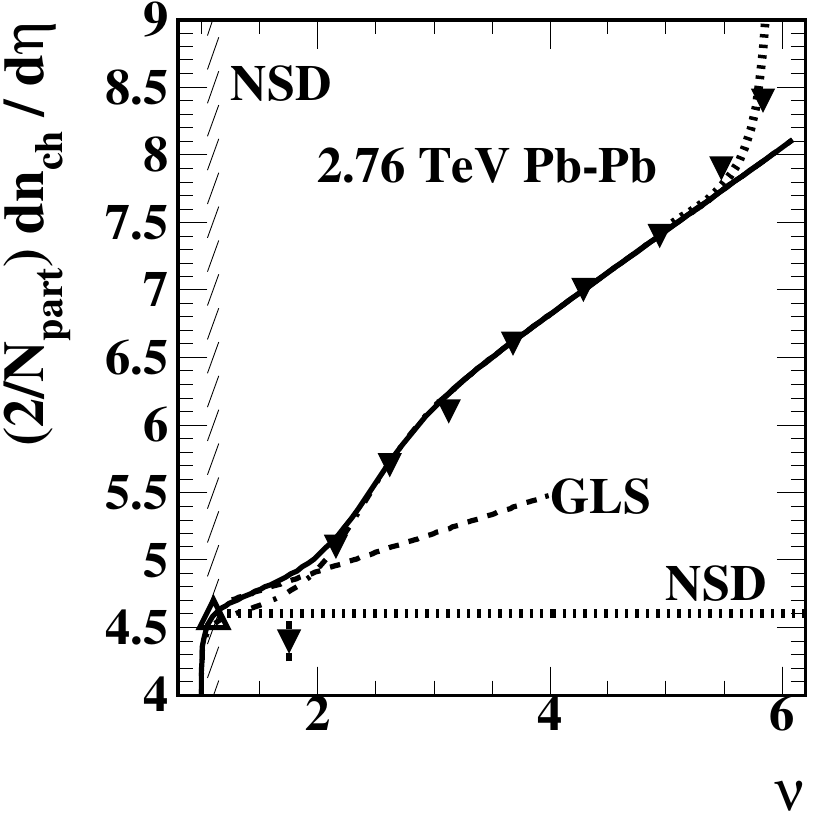}
\put(-20,46) {\bf (a)}
  \includegraphics[height=.25\textwidth]{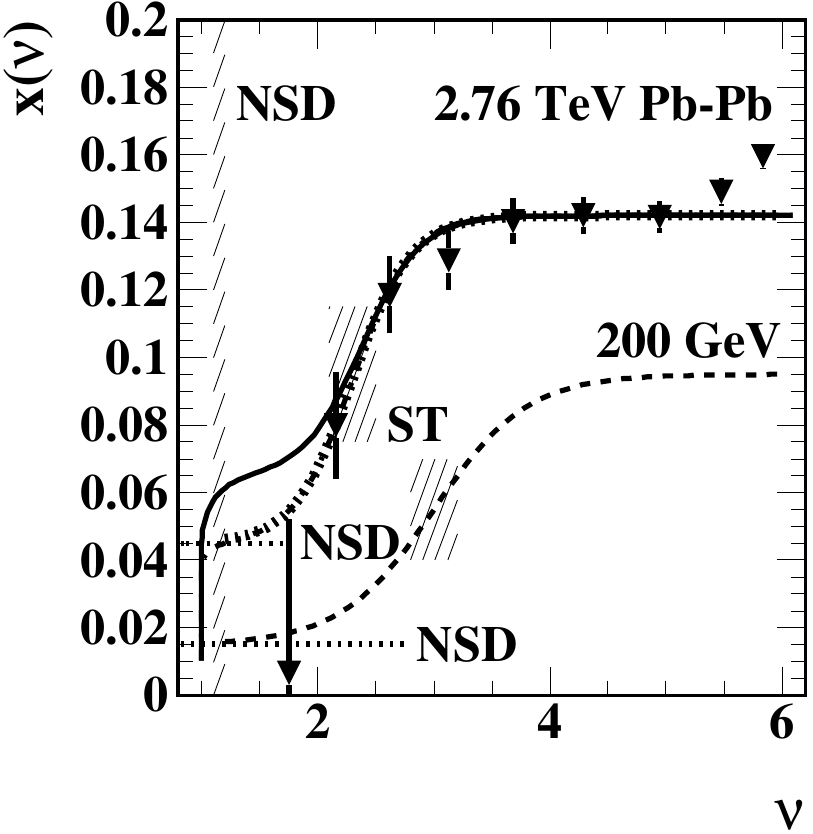}
\put(-20,40) {\bf (b)}
  \includegraphics[height=.25\textwidth]{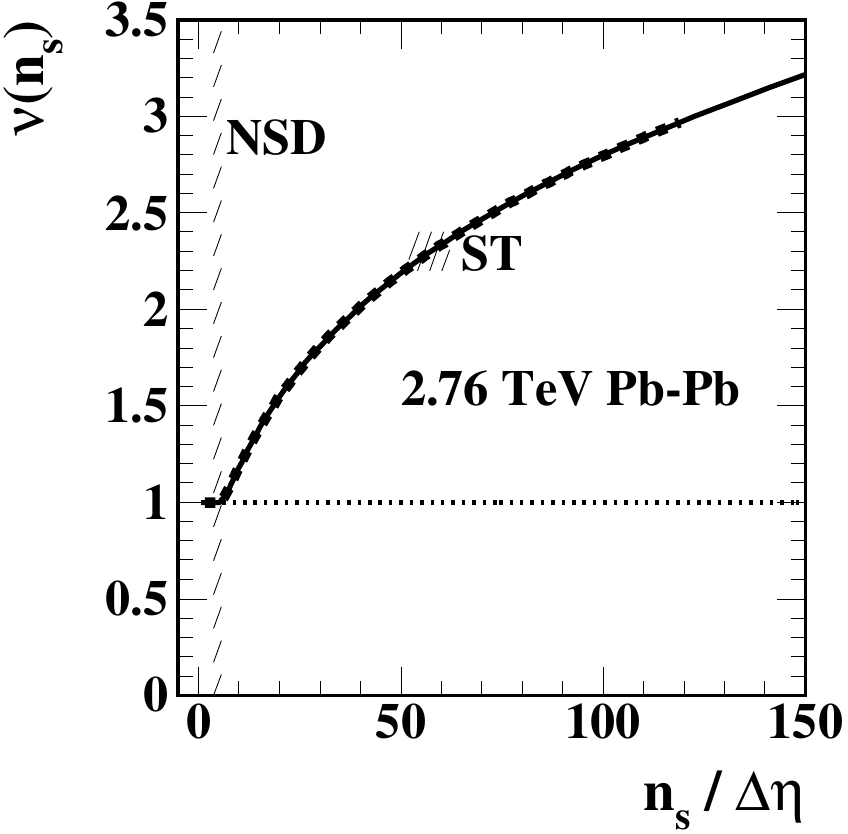}
\put(-24,65) {\bf (c)}
  \includegraphics[height=.25\textwidth]{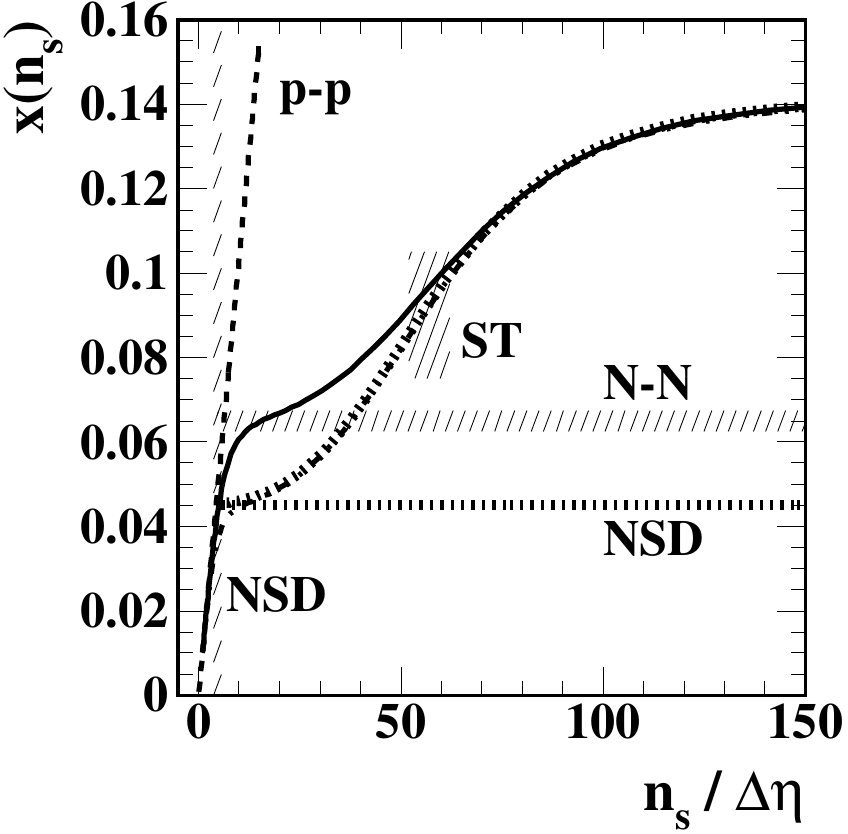}
\put(-24,65) {\bf (d)}
\caption{\label{pbpbmpt}
(a) Per-participant-pair charge density vs centrality measure $\nu$ (points) from Ref.~\cite{alicemultpbpb} compared to the TCM (solid curve).
(b) Data from panel (a) transformed to TCM parameter $x(\nu)$ compared to the TCM of Eq.~(\ref{xnu}) (solid, bold-dotted curves).
(c) Glauber parameter $\nu(n_s)$ defined in the text.
(d) Parameter $x(\nu)$ from panel (b) transformed to $x(n_s)$ based on $\nu(n_s)$ in panel (c).
} 
 \end{figure}

Figure~\ref{pbpbmpt} (a) shows per-participant-pair yield data from Ref.~\cite{alicemultpbpb} (points). The dashed line is a Glauber linear superposition (GLS) model representing no jet modification. The upper bold dotted curve shows the expected effect of multiplicity fluctuations for more-central \aa\ collisions~\cite{tomphenix}. The solid curve is the TCM in Eq.~(\ref{nchtcm}) with $x(\nu)$ defined below. Panel (b) shows solution of Eq.~(\ref{nchtcm}) for parameter $x(\nu)$ (points) assuming $\bar \rho_{sNN} \rightarrow \bar \rho_{sNSD} \approx 4.6$~\cite{alicetomspec}. A {\em sharp transition} (ST), first observed for \auau\ jet-related 2D angular correlations~\cite{anomalous}, representing evidence of jet modification in the yield trend~\cite{fragevo} is evident. The bold dotted curve is a simple TCM expression for $x(\nu)$ that starts from the NSD \pp\ value $x_{pp} \approx 0.045$, increases rapidly through the ST and maintains a saturation value $\approx 0.142$ for more-central collisions, approximately threefold increase over the NSD value. The dashed curve is the equivalent for 200 GeV \auau\ collisions. The interval $\bar \rho_s = 10$ - 40 is not well-defined {\em a priori} by yield data so a TCM expression is required that includes an adjustable $x_{pp}$~\cite{alicetommpt}
\bea \label{xnu}
x(\nu) = x_{pp} + (0.142 - x_{pp}) \left\{  1 + \tanh[(\nu - 2.3) / 0.5] \right\}/2,
\eea
with ST near $\nu = 2.3$.

Figure~\ref{pbpbmpt} (c) shows $\nu(n_s) \approx ( \bar \rho_{s} / \bar \rho_{sNSD})^{1/3}$ with lower bound $\nu = 1$. Panel (d) shows $x(n_s) = x[\nu(n_s)]$ based on Eq.~(\ref{xnu}) and panel (c). For this model peripheral \pbpb\ follows the  \pp\ trend (dashed line) for the most peripheral collisions, and more-central \pbpb\ shows the ST feature (jet modification). Yield data do not indicate where the transition from noneikonal \pp\ to eikonal Glauber trend occurs, requiring the adjustable $x_{pp}$ parameter in Eq.~(\ref{xnu}).

\begin{figure}[h]
\centering
  \includegraphics[height=.25\textwidth]{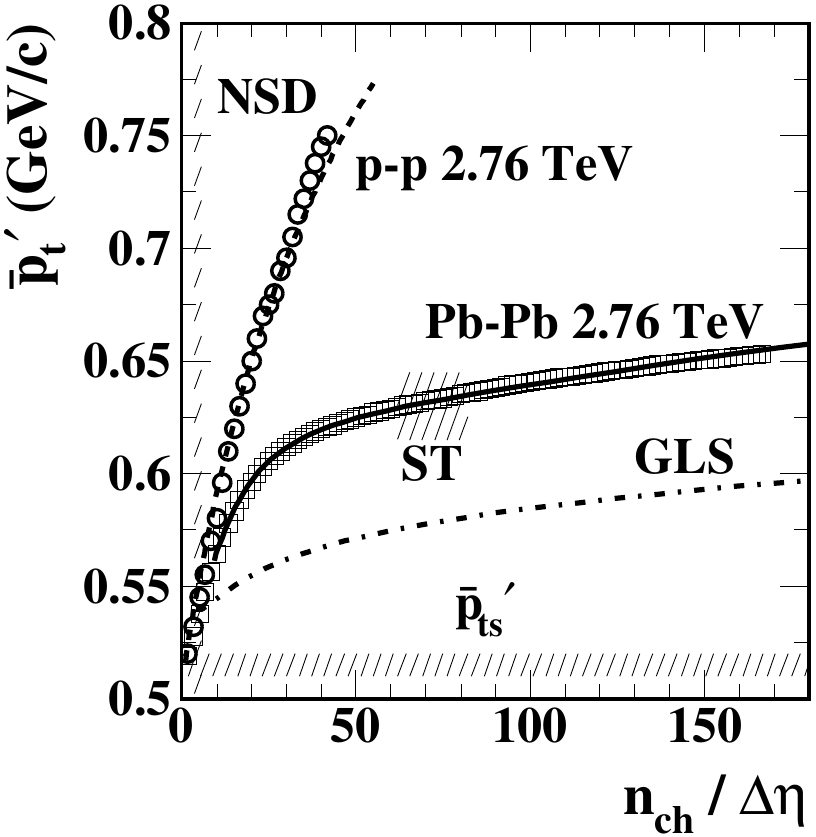}
\put(-20,80) {\bf (a)}
  \includegraphics[height=.25\textwidth]{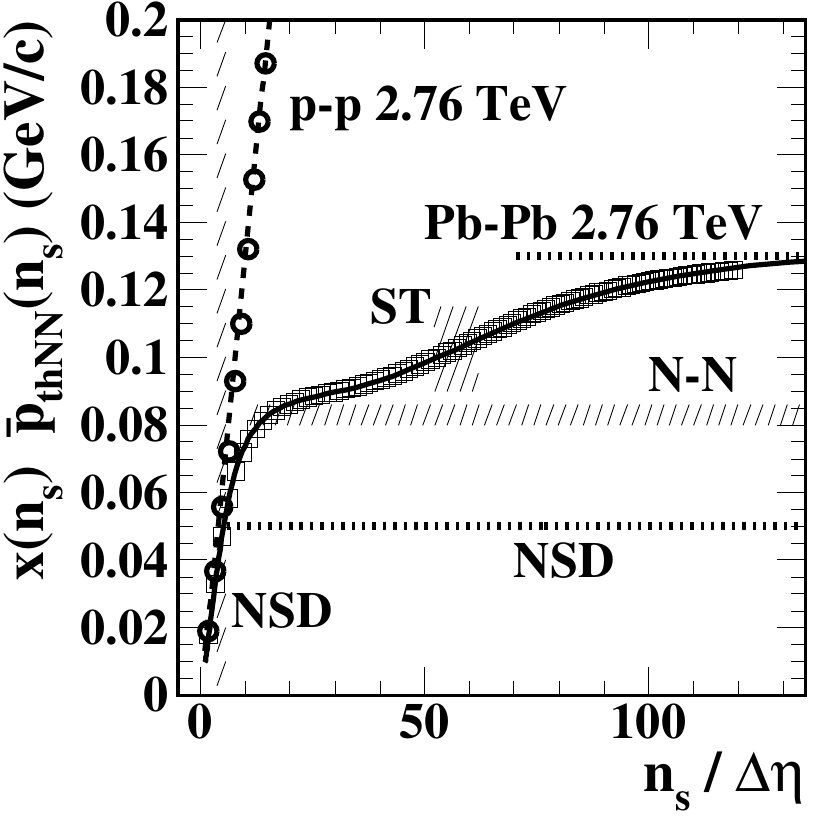}
\put(-20,80) {\bf (b)}
  \includegraphics[height=.25\textwidth]{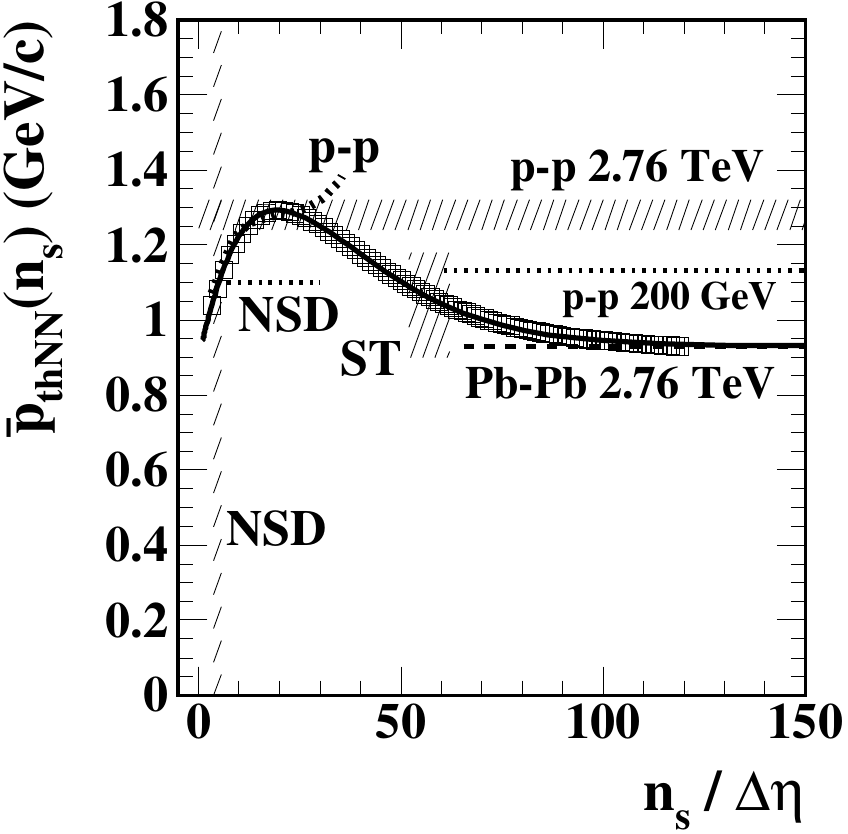}
\put(-24,30) {\bf (c)}
  \includegraphics[height=.25\textwidth]{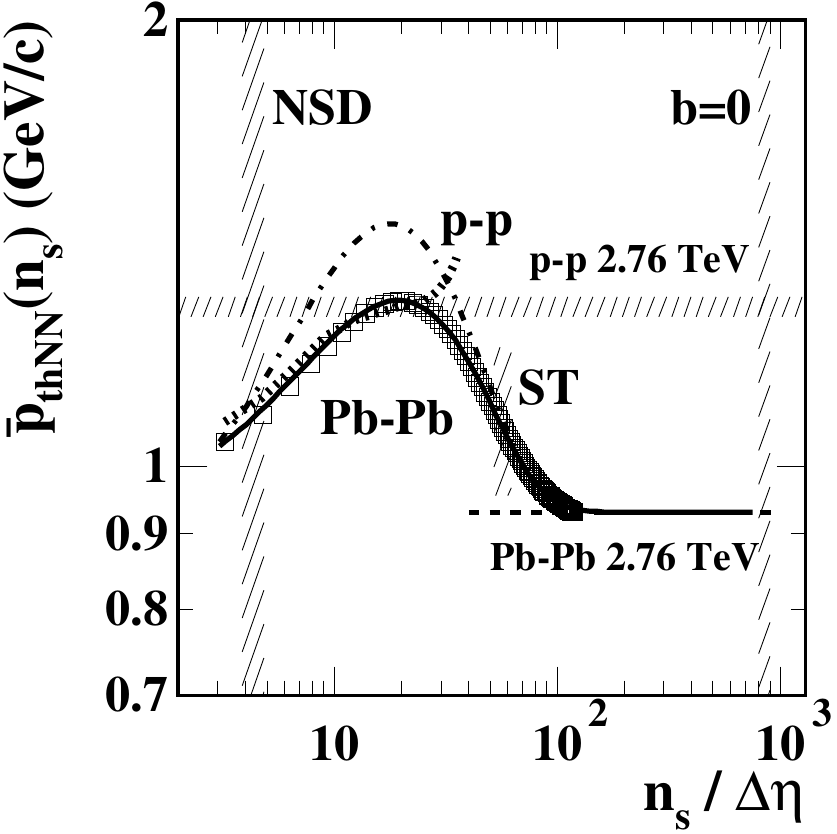}
\put(-64,30) {\bf (d)}
\caption{\label{pbpbdata}
(a) Uncorrected $\bar p_{t}'$ data (points) from Ref.~\cite{alicempt} vs the \mmpt\ TCM (solid curve). \pp\ data (open circles) and TCM (dashed curve) are included for comparison. The GLS curve represents the \pbpb\ trend for no jet modification.
(b) Product $x(n_s) \, \bar p_{thNN}(n_s)$ obtained from the data in panel (a) by solving Eq.~(\ref{ppbtcm}) using $\nu(n_s)$ in Fig.~\ref{pbpbmpt} (c).
(c) $\bar p_{thNN}(n_s)$ obtained from data in panel (b) using $x(n_s)$ in Fig.~\ref{pbpbmpt} (d).
(d) Panel (c) plotted in log-log format emphasizing the most-peripheral interval.
} 
 \end{figure}

Figure~\ref{pbpbdata} (a) shows uncorrected $\bar p_t'$ data for 2.76 TeV \pbpb\ collisions from Ref.~\cite{alicempt} (open squares). The uncorrected soft component is $\bar p_{ts}' \approx 0.51$ GeV/c. Data for \pp\ collisions are also shown (open circles). The \pbpb\ \mmpt\ TCM (solid curve) follows the \ppb\ TCM in Eq.~(\ref{ppbtcm}) (first) but with the $\nu(n_s)$ and $x(n_s)$ models in Fig.~\ref{pbpbmpt} (c,d). The dashed curve is the corresponding TCM for \pp\ data. The dash-dotted curve is a GLS model assuming no jet modification.

Figure~\ref{pbpbdata} (b) shows  product $x(n_s) \bar p_{thNN}(n_s) \approx \bar P_{thNN} / n_{sNN}$ (open squares) obtained from the data in panel (a) according to  Eqs.~(\ref{ppbtcm}) given the $\nu(n_s)$ model in Fig.~\ref{pbpbmpt} (c). The open circles are corresponding \pp\ data. The solid curve is the \pbpb\ TCM and the dashed curve is the \pp\ TCM. Note that the ST is clearly evident in this more-differential plot format, but not in panel (a). The \pbpb\ data follow the \pp\ trend closely up to and even beyond the NSD $\bar \rho_s$ value (vertical hatched band). 

Figure~\ref{pbpbdata} (c) shows  $\bar p_{thNN}(n_s)$ data (points) obtained from panel (b) using the $x(n_s)$ model from Fig.~\ref{pbpbmpt} (d). The inferred $\bar p_{thNN}(n_s)$ values follow the \pp\ trend (bold dotted curve just visible) for peripheral \pbpb\ but then descend through  the ST to a saturation value that is about 75\% of the maximum. The constant saturation values for $x(n_s)$ and $\bar p_{thNN}(n_s)$ extend from just above the ST ($\bar \rho_s \approx 100$) to central collisions ($\bar \rho_s \approx 850$) suggesting that jets do not undergo further alteration within that interval.
Panel (d) shows the same features in a log-log format providing better access to peripheral details and extending the plot to central \pbpb\ collisions. The correspondence between \pp\ (dotted curve) and \pbpb\ data (open squares) up to $\bar \rho_s \approx 20$ ($4 \times \bar \rho_{sNSD}$) is evident. The dash-dotted curve marks where the data would lie if the bold dotted curve in Fig.~\ref{pbpbmpt} (d) were used for $x(n_s)$. The data were actually  transformed using the solid curve in that panel with $x_{pp} \rightarrow 0.065$ in Eq.~(\ref{xnu}), 1.45 times the NSD value. The solid curve  through the $\bar p_{thNN}(n_s)$ data, representing new information obtained from \pbpb\ \mmpt\ data, is a parametrization that forms the third element of the \pbpb\ \mmpt\ TCM~\cite{alicetommpt}.


\section{Comment on Glauber-model analysis of p-Pb collision centrality} \label{}

The TCM for \ppb\ \mmpt\ data described in Sec.~\ref{ppb} and Ref.~\cite{alicetommpt} can be compared with a recent Glauber analysis of 5 TeV \ppb\ centrality in Ref.~\cite{aliceppbprod}. For the Glauber analysis it is assumed that integrated charge multiplicity is proportional to Glauber $N_{part}$: ``[$n_{ch}$] at mid-rapidity scales linearly with [$N_{part}$].'' That assumption is equivalent to assuming that a probability (i.e.\ event-frequency) distribution on charge multiplicity -- $dP/dn_{ch}$ -- is equivalent to a fractional cross-section distribution on $N_{part}$ -- $(1/\sigma_0)d\sigma / dN_{part}$ -- inferred from a Glauber Monte Carlo.

\begin{figure}[h]
\centering
  \includegraphics[height=.235\textwidth,width=.28\textwidth]{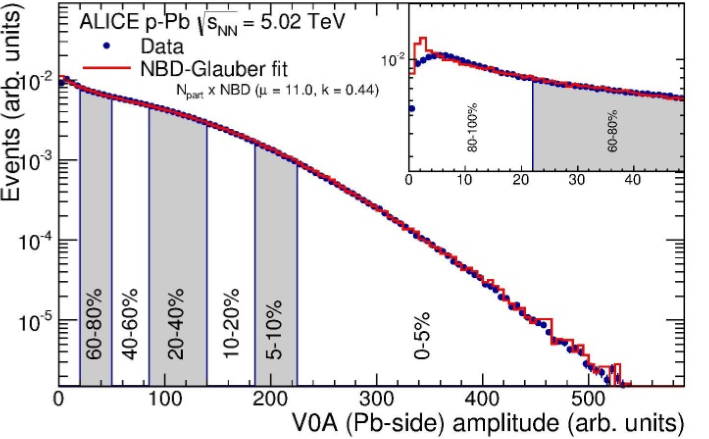}
\put(-23,40) {\bf (a)}
  \includegraphics[height=.24\textwidth]{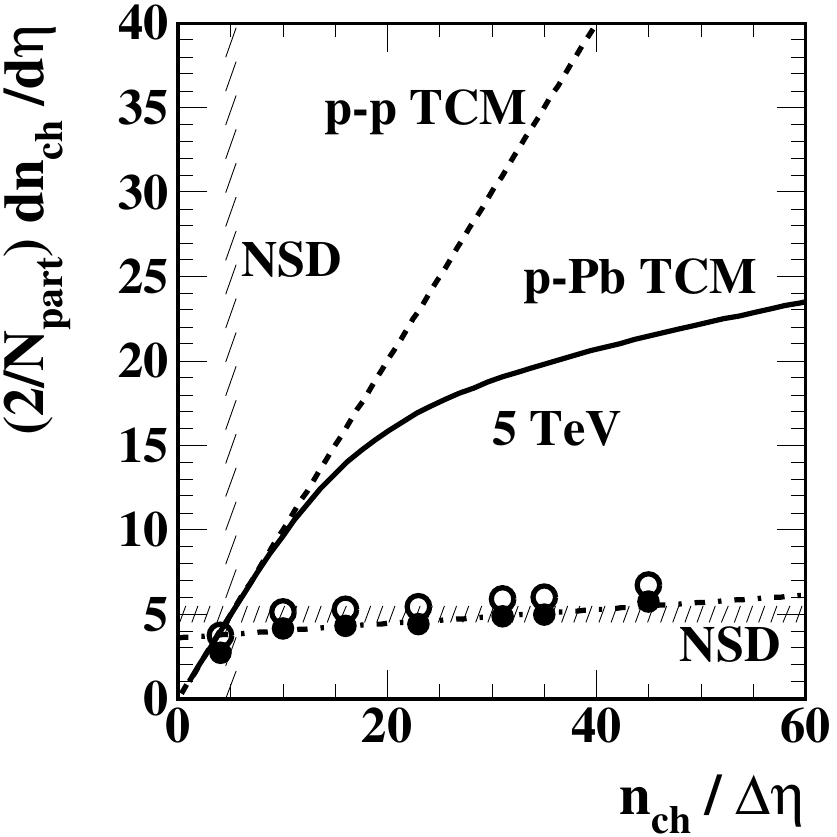}
\put(-23,40) {\bf (b)}
  \includegraphics[height=.24\textwidth]{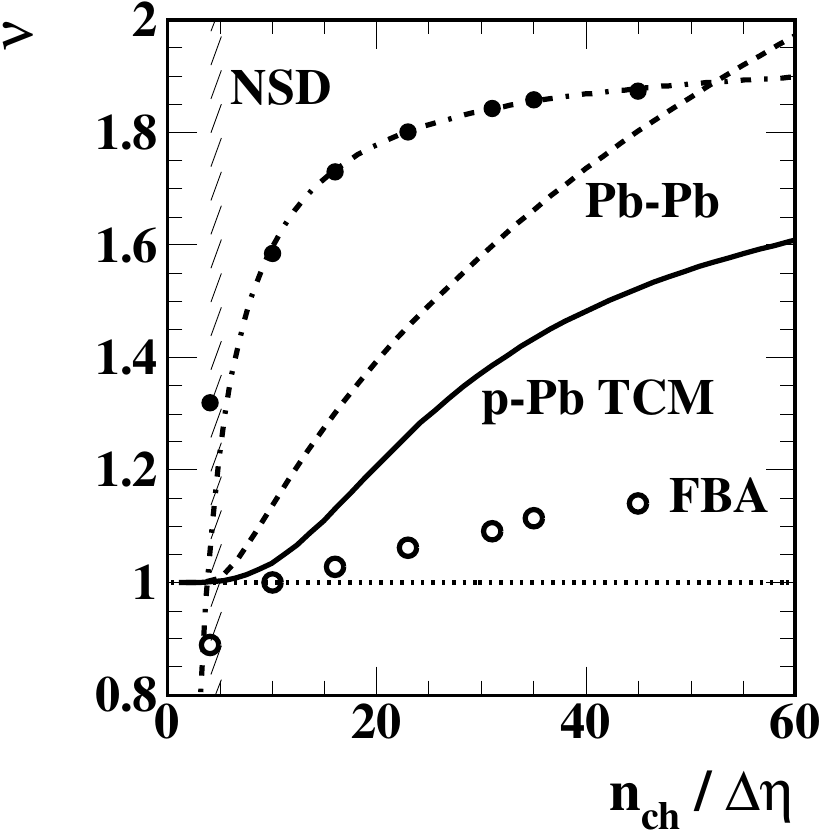}
\put(-20,20) {\bf (c)}
  \includegraphics[height=.24\textwidth]{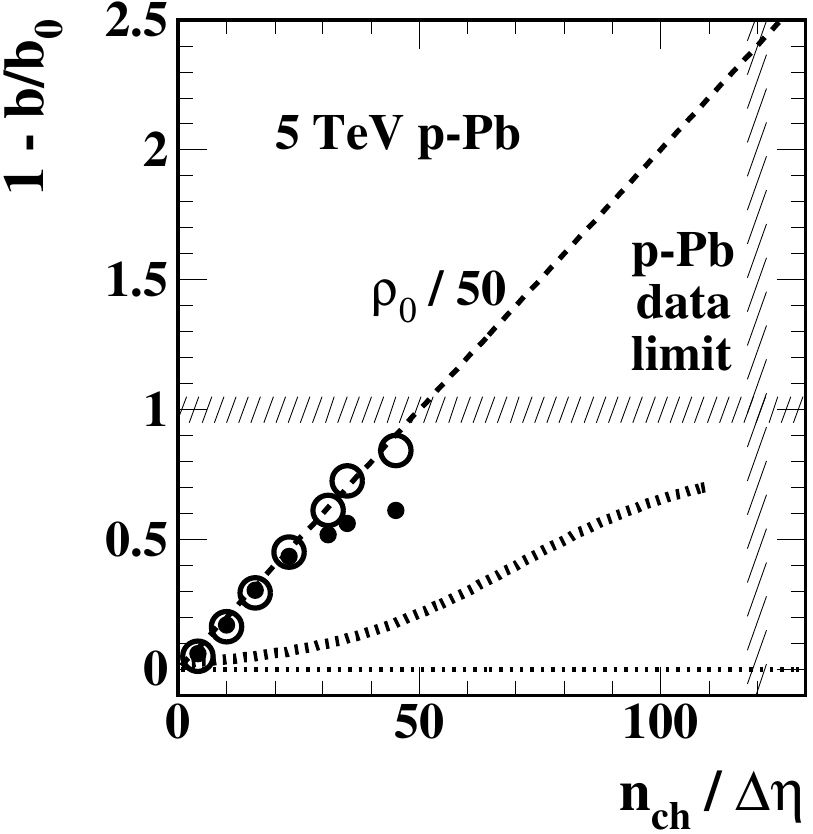}
\put(-24,24) {\bf (d)}
\caption{\label{v0a}
(a) Event-frequency distribution on charge multiplicity in a VOA detector with centrality intervals assigned by \ppb\ Glauber analysis~\cite{aliceppbprod}.
(b) Per-participant-pair \ppb\ yields inferred from the Glauber analysis (solid points) and from the present study (solid curve).
(c) Participant path length $\nu$ inferred from the Glauber analysis (solid points) and the present study (solid curve). \pbpb\ Glauber results (dashed) are included for comparison.
(d) Fractional impact parameter in the form $1 - b/b_0$ inferred from the Glauber study (solid points) and indirectly from fractional cross sections in panel (a) (open circles) compared to a trend suggested by the present study (bold dotted curve).
} 
 \end{figure}

Figure~\ref{v0a} (a) shows an event-frequency distribution (histogram) on VOA amplitude (charge multiplicity within a pseudorapidity interval in the Pb hemisphere) from Ref.~\cite{aliceppbprod}. The assumption $N_{part} \propto n_{ch}$ is equivalent to $dP/dn_{ch} \sim (1/\sigma_0)d\sigma / dN_{part}$ as noted above, and the fractional cross sections in the form $100(1 - \sigma / \sigma_0)$ noted in the panel and apparently obtained from a running integral of the histogram seem to be consistent with that statement.

Figure~\ref{v0a} (b) shows Glauber estimates of $(2/N_{part}) dn_{ch}/d\eta$ from Ref.~\cite{aliceppbprod} (solid points) compared to the TCM equivalent for \pp\ (dashed) and \ppb\ (solid) collisions from Ref.~\cite{alicetommpt} as reported above. The large discrepancy is apparent. Given $N_{part}$ and $N_{bin}$ from the Glauber analysis corresponding values of $\nu$ are plotted in panel (c) (solid dots). The TMC equivalent is the solid curve, and the Glauber trend for \pbpb\ collisions is shown as the dashed curve. That $\nu$ for \ppb\ collisions might anywhere greatly exceed that for \pbpb\ collisions is notable.

Figure~\ref{v0a} (d) shows fractional impact parameter in the form $1 - b/b_0$ (open circles) inferred from the fractional cross sections in panel (a) assuming that $\sigma / \sigma_0 = (b / b_0)^2$. The solid dots are the same quantity based on $b$ values from the Glauber analysis and assuming that $b_0 \approx 7.1 + 0.85 \approx 8$ fm. Whereas the Glauber analysis in Ref.~\cite{aliceppbprod} reports 0-5\% central \ppb\ collisions for $\bar \rho_0 \approx 45$ the \ppb\ \mmpt\ data from Ref.~\cite{alicempt} (same collaboration) extend to $\bar \rho_0 \approx 120$, nearly three times larger. The bold dotted curve suggests a more realistic trend for $b/b_0$.

The basic problem with the analysis in Ref.~\cite{aliceppbprod} is the initial assumption $N_{part} \propto n_{ch}$ which is strongly violated for \ppb\ collisions. Just as for \pp\ collisions in Sec.~\ref{pp} initial increase of the \nch\ condition is more easily met in \ppb\ collisions by higher multiplicities in individual \pn\ collisions than by increasing \ppb\ centrality. $N_{part}/2 \approx 1$ persists until $\bar \rho_{s0} \approx 15$, three times the NSD value. Only then does $N_{part} / 2$ begin to increase above 1 significantly. Below that point $d\sigma / dn_{ch} \approx 0$ and \ppb\ ``centrality'' remains fixed at 100\%. The correct relation is
\bea
d\sigma/dn_{ch} &=& (dN_{part}/dn_{ch}) d\sigma / dN_{part},
\eea
where $d\sigma / dN_{part}$ may be obtained from a Glauber Monte Carlo but the Jacobian $dN_{part}/dn_{ch}$ is determined by the \ppb\ \mmpt\ TCM as described above. As noted, the Jacobian factor is essentially zero over a substantial \nch\ interval below transition point $\bar \rho_{s0}$ in Sec.~\ref{ppb}. Based on the assumption in Ref.~\cite{aliceppbprod} the Jacobian would have a fixed value near $2/n_{chNSD}$ for all \nch.


\section{Summary} \label{summ}

Recently, claims have emerged that hydrodynamic flows must play a role in smaller collision systems (\pa\ and even \pp) based on observation of the same phenomena in those systems attributed to flows in more-central \aa\ systems. However, the argument could be reversed to conclude that if phenomena appear in collision systems so small and low-density that flows arising from particle rescattering are impossible then the same phenomena observed in more-central \aa\ may not represent flows. Recent \mmpt\ data may contribute to a resolution.

The two-component (soft + hard) model (TCM) of hadron production near midrapidity provides an accurate and comprehensive description of yield, spectrum and correlation data applicable to any A-B collision system. While one or more other mechanisms may contribute to hadron production in minor ways (e.g.\ a nonjet quadrupole component) projectile-nucleon dissociation (soft) and dijet production (hard) appear as the dominant processes in all cases.

From TCM analysis of \mmpt\ data described above for three successive collision systems the following may be concluded: (a) \pp\ $\bar p_{th}(n_s,\sqrt{s})$ hard-component trends agree with spectrum HC evolution and measured MB dijet properties. (b) \pp\ dijet production is {\em noneikonal}, \pp\ centrality is not relevant. (c) The \ppb\ $\bar p_{th}(n_s)$ hard component establishes factorization of A-B Glauber and \nn\ noneikonal trends. (d) \ppb\ \mmpt\ data confirm that MB dijets dominate $\bar p_{t}(n_s)$ variation. (e) In effect, MB dijets probe the centrality evolution of \ppb\ collisions with \nch. (f) \pbpb\ \mmpt\ data confirm that the Glauber model effectively describes more-central \aa\ collisions, but peripheral collisions accurately follow \pp\ (i.e.\ \nn) trends. (g) The \pbpb\ $\bar p_{thNN}(n_s)$ trend confirms that jets are {\em modified} quantitatively above the ST, but jets still dominate spectrum and correlation structure in more-central \pbpb\ collisions. 

Thus, the likelihood that flows play any role in small collision systems appears negligible. Claimed novel phenomena may be better explained either as due to MB dijets or as a related QCD process (e.g.\ multipole radiation). The same argument may be extended even to the most-central \aa\ collisions at the highest energies. Before novelty is claimed the contribution of MB dijets and related QCD processes to data trends must first be well understood.

\end{document}